\documentclass[twocolumn,Astrosymb]{aastex62}
\graphicspath{{./}{figures/}}

\received{December 27, 2019}
\revised{June 6, 2020}
\accepted{June 12, 2020}
\submitjournal{PASP}

\shorttitle{Multiband Photometry Simulations}
\shortauthors{Louie et al.}

\begin{document}

\title{Simulations Predicting the Ability of Multi-Color Simultaneous Photometry to \\ Distinguish TESS Candidate Exoplanets from False Positives}

\correspondingauthor{Dana R. Louie}
\email{danalouie@astro.umd.edu}

\author[0000-0002-2457-272X]{Dana R. Louie}
\affiliation{Department of Astronomy, University of Maryland, College Park, MD 20742, USA}

\author[0000-0001-8511-2981]{Norio Narita}
\affiliation{Astrobiology Center, National Institutes of Natural Sciences, 2-21-1 Osawa, Mitaka, Tokyo 181-8588, Japan}
\affiliation{JST, PRESTO, 2-21-1 Osawa, Mitaka, Tokyo 181-8588, Japan}
\affiliation{National Astronomical Observatory of Japan, National Institutes of Natural Sciences, 2-21-1 Osawa, Mitaka, Tokyo 181-8588, Japan}
\affiliation{Instituto de Astrof\'isica de Canarias, V\'ia L\'actea s/n, E-38205 La Laguna, Tenerife, Spain
}

\author{Akihiko Fukui}
\affiliation{Department of Earth and Planetary Science, Graduate School of Science, The University of Tokyo, 7-3-1 Hongo, Bunkyo-ku, Tokyo 113-0033, Japan
}
\affiliation{Instituto de Astrof\'isica de Canarias, V\'ia L\'actea s/n, E-38205 La Laguna, Tenerife, Spain
}

\author[0000-0003-0987-1593]{Enric Palle}
\affiliation{Instituto de Astrof\'isica de Canarias, V\'ia L\'actea s/n, E-38205 La Laguna, Tenerife, Spain
}
\affiliation{Departamento de Astrof\'isica, Universidad de La Laguna (ULL), 38206, La Laguna, Tenerife, Spain}

\author{Motohide Tamura}
\affiliation{Graduate School of Science, The University of Tokyo, Japan}
\affiliation{Astrobiology Center, National Institutes of Natural Sciences, 2-21-1 Osawa, Mitaka, Tokyo 181-8588, Japan}
\affiliation{National Astronomical Observatory of Japan, National Institutes of Natural Sciences, 2-21-1 Osawa, Mitaka, Tokyo 181-8588, Japan}

\author[0000-0001-9194-1268]{Nobuhiko Kusakabe}
\affiliation{Astrobiology Center, National Institutes of Natural Sciences, 2-21-1 Osawa, Mitaka, Tokyo 181-8588, Japan}

\author{Hannu Parviainen}
\affiliation{Instituto de Astrof\'isica de Canarias, V\'ia L\'actea s/n, E-38205 La Laguna, Tenerife, Spain
}
\affiliation{Departamento de Astrof\'isica, Universidad de La Laguna (ULL), 38206, La Laguna, Tenerife, Spain}


\author{Drake Deming}
\affiliation{Department of Astronomy, University of Maryland, College Park, MD 20742, USA}



\begin{abstract}

The Transiting Exoplanet Survey Satellite (TESS) is currently concluding its 2-year primary science mission searching 85\% of the sky for transiting exoplanets. TESS has already discovered well over one thousand TESS objects of interest (TOIs), but these candidate exoplanets must be distinguished from astrophysical false positives using other instruments or techniques. The 3-band Multi-color Simultaneous Camera for Studying Atmospheres of Transiting Planets (MuSCAT), as well as the 4-band MuSCAT2, can be used to validate TESS discoveries. Transits of exoplanets are achromatic when observed in multiple bandpasses, while transit depths for false positives often vary with wavelength.  We created software tools to simulate MuSCAT/MuSCAT2 TESS follow-up observations and reveal which planet candidates can be efficiently distinguished from blended eclipsing binary (BEB) false positives using these two instruments, and which must be validated using other techniques.  We applied our software code to the Barclay et al. (2018) predicted TESS discoveries, as well as to TOIs downloaded from the ExoFOP-TESS website.  We estimate that MuSCAT (MuSCAT2 values in parentheses) will be able to use its multi-color capabilities to distinguish BEB false positives for $\sim$17\% ($\sim$18\%) of all TESS discoveries, and $\sim$13\% ($\sim$15\%) of $R_{\rm pl} < 4R_\oplus$ discoveries.  Our TOI analysis shows that MuSCAT (MuSCAT2) can distinguish BEB false positives for $\sim$55\% ($\sim$52\%) of TOIs with transit depths greater than 0.001, for $\sim$64\% ($\sim$61\%) of TOIs with transit depths greater than 0.002, and for $\sim$70\% ($\sim$68\%) of TOIs with transit depth greater than 0.003.  Our work shows that MuSCAT and MuSCAT2 can validate hundreds of $R_{\rm pl} < 4R_\oplus$ candidate exoplanets, thus supporting the TESS mission in achieving its Level 1 Science Requirement of measuring the masses of 50 exoplanets smaller in size than Neptune.  Our software tools will assist scientists as they prioritize and optimize follow-up observations of TESS objects of interest.   
\end{abstract}



\section{Introduction} \label{sec:intro}

The Transiting Exoplanet Survey Satellite (TESS), which launched 18 April 2018, is projected to detect over one thousand transiting exoplanets smaller than Neptune \citep{2015ApJ...809...77S,2016SPIE.9904E..2BR,2018ApJS..239....2B}.  However, \citet{2015ApJ...809...77S}  showed that TESS will also detect several thousand astrophysical false positives, produced by blended light from a target star and eclipsing binary stars in the foreground/background, or bound to the target star.  The TESS Follow-Up Observing Program (TFOP)\footnote{\url{https://tess.mit.edu/followup/}} will facilitate achievement of the TESS Level 1 Science Requirement to measure the masses of 50 exoplanets smaller in size than Neptune.  The first step in the TFOP pipeline is to validate candidate exoplanets by distinguishing true exoplanets from astrophysical false positives. 

Transiting exoplanets can be distinguished from astrophysical false positives by determining the wavelength/color-dependence of the amount of stellar light received--transiting exoplanets are largely achromatic when observed in different bandpasses \citep{2004ESASP.538..255A,2019A&A...630A..89P}.  Ground-based multiband photometry makes use of this color-dependence to distinguish true exoplanets from astrophysical false positives.  The Multi-color Simultaneous Camera for Studying Atmospheres of Transiting Planets, or MuSCAT \citep{2015JATIS...1d5001N}, a 3-color multiband photometer used on the National Astronomical Observatory of Japan (NAOJ) 1.88-m telescope at Okayama Astro-Complex (OAC), Japan, is one instrument the exoplanet community uses for TESS validation.  The 4-color MuSCAT2 instrument \citep{2019JATIS...5a5001N} installed on the 1.52-m Carlos Sanchez Telescope at Teide Observatory in the Canary Islands also supports TESS validations with more than 200 dedicated observing nights per year.\footnote{http://vivaldi.ll.iac.es/OOCC/iac-managed-telescopes/telescopio-carlos-sanchez/muscat2/} 

\citet{2015ApJ...809...77S} analyzed the types of false positives that TESS would detect when observing 200,000 preselected target stars at 2-minute cadences, and they found that TESS would discover 1103 $\pm$ 33 eclipsing binary (EB) systems which fall into the following three categories:\footnote{Additional false positives would be discovered from full-frame image data, but \citet{2015ApJ...809...77S} only analyzed the 2-minute cadence data.}  
        
\begin{enumerate}
\item Eclipsing binaries (EB): the target star is part of a binary system, and it is grazed by eclipses from its companion.  For example, a solar type star may be grazed by eclipses from a late M dwarf companion.
\item Hierarchical EB (HEB): the target star is a triple or higher-order system, and one pair of stars eclipses.  For example, if the target star is a solar type star, and another solar type star in the system is eclipsed by an M dwarf, the light from the target star will dilute the eclipse depth of the EB, producing a light curve similar to that of a planetary transit of the target star.
\item Blended EB (BEB): the target star blends with an EB in the foreground/background within the photometric aperture of the target star.  This case is similar to an HEB, except that here the EB is not gravitationally bound to the target star.  Note this may also be referred to as a nearby eclipsing binary (NEB).
\end{enumerate}

\citet{2015ApJ...809...77S} examined the possibility of using TESS observation data to distinguish false positives through (1) ellipsoidal variations, (2) secondary eclipses, (3) lengthy ingress and egress durations, or (4) centroid motion of the image on the detector.  They found that these methods provide clues to help distinguish EBs for 98.6\% of the EBs and 93.0\% of the HEBs.  However, roughly one-quarter of the BEBs could not be distinguished from exoplanets using any of these methods, leaving $\sim$150 of the false positives indistinguishable from actual exoplanets.  

For those cases where TESS observation data cannot distinguish transiting exoplanets from false positives, we turn to ground-based multiband photometry.  When a star is observed in a given photometric bandpass, both transiting exoplanets and eclipsing binary stars decrease the amount of light received as the planet transits, or as one star of a binary pair eclipses the other star.  Transiting exoplanets block a portion of the host star's light, and thus the decreases in host starlight received do not depend significantly on wavelength.  False positives caused by blended light from target stars and eclipsing binaries also produce signals as one star eclipses its binary, but as long as the colors of the stars are significantly different, the signal is much more wavelength dependent \citep[see, e.g.,][]{2012MNRAS.426..342C}.  This wavelength dependence would be evident by comparing the amount of light received in MuSCAT's three bandpasses (or in MuSCAT2's four bandpasses), which collect light over different wavelengths.  Notably, the sensitivity of the instrument to a given signal will depend on the physical parameters (e.g. radius, temperature, etc.) of the star(s) and planet being observed.  These physical parameters will vary tremendously between TESS planet candidates.  

Our goal is to simulate MuSCAT and MuSCAT2 follow-up observations to reveal which planet candidates can be efficiently distinguished from BEB false positives using these instruments.  This understanding will allow TFOP working group members to better prioritize and optimize follow-up validations of TESS detections.  We also examine our results to determine any common characteristics between those planet candidates that can be validated using MuSCAT and MuSCAT2.  In addition, we provide a software tool to assist in planning MuSCAT and MuSCAT2 follow-up observations.  The code reads in a list of parameters for several TESS discoveries (e.g. a list of TESS objects of interest (TOIs) released to the community on the Exoplanet Follow-up Observing Program for TESS (ExoFOP-TESS) website\footnote{\url{https://exofop.ipac.caltech.edu/tess/}}), and predicts the probability that each discovery can be successfully distinguished from BEB false positives using MuSCAT and/or MuSCAT2.

This work presents the results of a computationally simple analysis that can be easily applied to a large sample of TESS candidate exoplanets.  Recently, \citet{2019A&A...630A..89P} showed that multi-color transit photometry can be used to determine the true radius ratio of an exoplanet candidate to its host star, when the light from the host star blends with unresolved light sources in the photometric aperture.  \citet{2019A&A...630A..89P}'s work is complementary to our current study since their analysis methods can be applied to any exoplanet candidate after actual observations are taken with MuSCAT or MuSCAT2.

Our paper is organized as follows.  In Section \ref{sec:methods}, we describe the simulated TESS exoplanet discoveries and the false positives we use in our analysis, and we explain the design and validation of our MuSCAT and MuSCAT2 simulation tools.  In Section \ref{sec:results}, we present our findings.  We summarize in Section \ref{sec:summary}.

\section{Methods} \label{sec:methods}

\begin{figure*}
\plotone{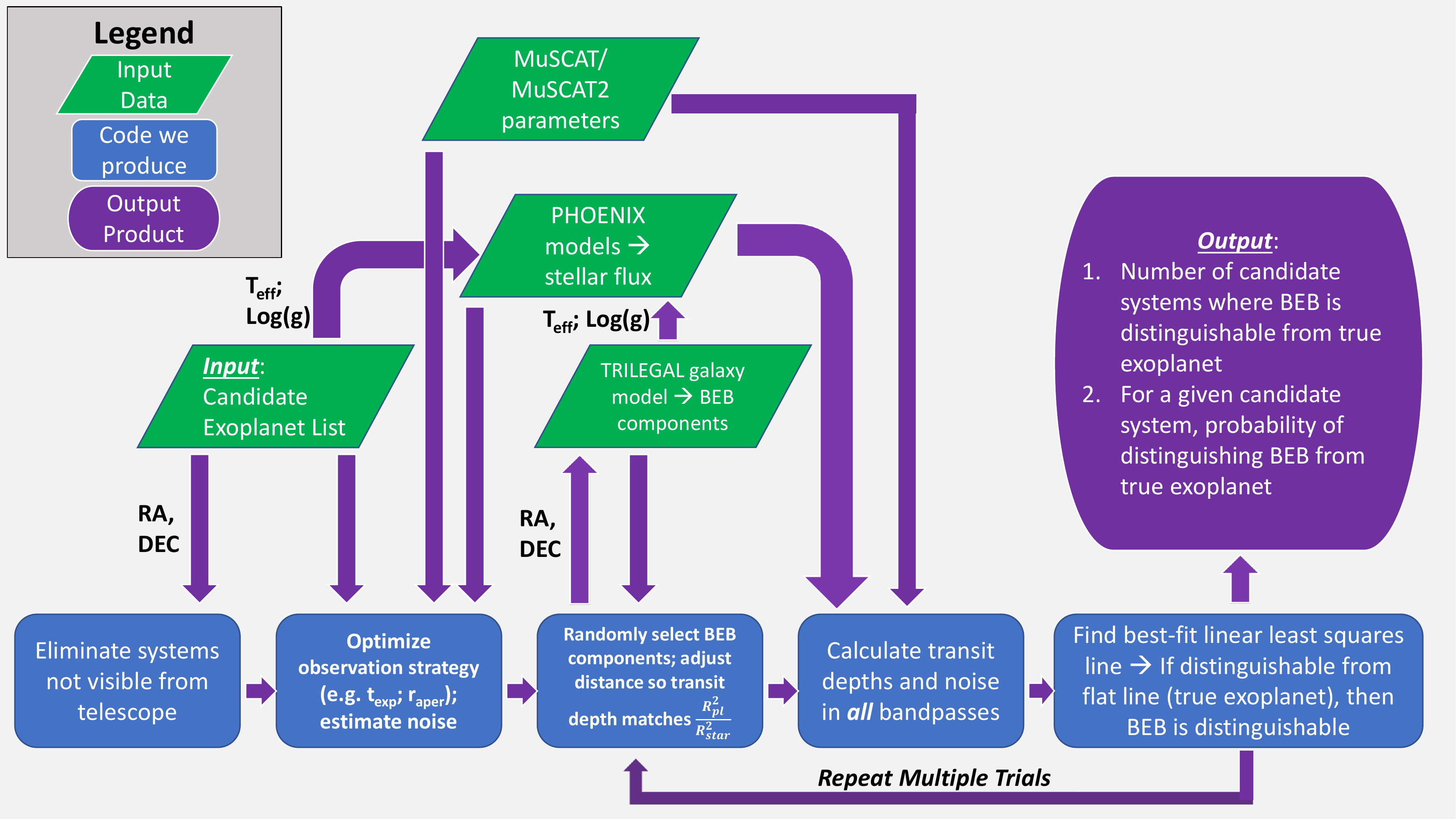}
\caption{Block diagram outlining the major components of our MuSCAT and MuSCAT2 simulation tools. \label{fig:SimBlockDiagram}}
\end{figure*}

True exoplanet transits are largely achromatic, but we expect false positives produced by blended eclipsing binary stars (BEBs) to produce different transit depths in the MuSCAT and MuSCAT2 bandpasses.  To predict MuSCAT and MuSCAT2 performance, we compare the transit depths we would obtain by observing a large sample of true exoplanets to the transit depths we would obtain in the MuSCAT and MuSCAT2 bandpasses by observing those same systems as BEB false positives.  We then determine the extent to which MuSCAT and MuSCAT2 can discriminate between true exoplanet and BEB transit depths.  We use the \cite{2018ApJS..239....2B} predicted TESS exoplanet discoveries (Section \ref{sec:ExoDisc}) for our candidate exoplanet sample.  In section \ref{sec:simtool}, we describe those aspects of the MuSCAT and MuSCAT2 instruments incorporated into our simulation tools.  We also describe our noise model.  We validated our simulation tools by comparing results produced by the simulations to those from observations of actual transiting exoplanets, incorporating a random factor to account for variations in quantities such as atmospheric transmittance and mirror reflectance (Section \ref{sec:validation}).  We calculate MuSCAT and MuSCAT2 transit depths for BEB false positives as explained in Section \ref{sec:FP}, and we determine the extent to which MuSCAT and MuSCAT2 can discriminate these BEBs using the criterion presented in Section \ref{sec:MuSCATcriterion}.  Finally, we applied our tools to TESS objects of interest (TOIs), as posted to the ExoFOP-TESS website.  Figure \ref{fig:SimBlockDiagram} shows a block diagram of our simulation routine.

\subsection{Predicted TESS Exoplanet Discoveries} \label{sec:ExoDisc}

\cite{2018ApJS..239....2B} used Monte Carlo methods to predict the properties of the exoplanets that TESS is likely to discover, and published a machine-readable file containing the properties of their 4,373 predicted planetary systems.\footnote{\url{https://iopscience.iop.org/article/10.3847/1538-4365/aae3e9/meta##apjsaae3e9t2}}  \citet{2018ApJS..239....2B} used stars in the TESS Input Catalog (TIC) Candidate Target List (CTL) \citep{2019AJ....158..138S}, employing Monte Carlo techniques to assign planets to the stars and determine how many of the planets TESS would detect.  They adopted Kepler planet occurrence rates from \cite{2013ApJ...766...81F} for AFGK stars, and from \cite{2015ApJ...807...45D} for M stars.  The predicted planet yield includes discoveries orbiting both pre-selected target stars viewed at 2-minute cadence, as well as stars viewed in full-frame images at 30-minute cadence.

Our simulations use the parameters in the \cite{2018ApJS..239....2B} machine-readable file as inputs.  For example, we determine whether a given system is observable from either Okayama Astro-Complex (OAC) or Teide Observatory using the system's right ascension and declination.  To calculate stellar flux emanating from the system, we use reported values of stellar effective temperature and log($g$) (calculated from radius and mass) to select an appropriate PHOENIX stellar model, and we then use reported TESS-band magnitude to scale the PHOENIX model.  We calculate transit depth using the radii of the planet and star.  Note that we ignore limb darkening in the transit depth calculation, but we deal with a large range of transit depths across our planetary candidates, and the effects of limb darkening are relatively small compared to other aspects of our simulation.

\subsection{MuSCAT and MuSCAT2 Simulation Tools} \label{sec:simtool}

Our simulation tools incorporate several components. First, we use PHOENIX stellar spectra to estimate the light received across each bandpass for a given star.  In addition, we include important details regarding MuSCAT\footnote{\url{http://esppro.mtk.nao.ac.jp/MuSCAT/observing.html}} \citep{2015JATIS...1d5001N} and MuSCAT2\footnote{\url{http://vivaldi.ll.iac.es/OOCC/iac-managed-telescopes/telescopio-carlos-sanchez/muscat2/}} \citep{2019JATIS...5a5001N} into our simulations, such as geographic location, telescope aperture, and throughput. We also make assumptions about observational selections such as exposure times and telescope defocusing based upon experience.  In this section, we describe each aspect of our code in detail, working from the targeted star to the instrument array in our description.

PHOENIX/BT-NextGen and PHOENIX/BT-Settl stellar emission spectrum grids \citep{2012RSPTA.370.2765A} provide stellar flux across the wavelength regimes of both the MuSCAT and MuSCAT2 instruments.  We employ PHOENIX stellar models for the target star, as well as the primary and secondary components of BEBs.  Our simulation routine selects the stellar model with effective temperature and $\log(g)$ values closest to those of the particular star(s) in the system we are modeling.  We employ solar metallicity spectra.  \cite{2018PASP..130d4401L} provide further details about the stellar models employed.

The stellar flux (ergs sec$^{-1}$ cm$^{-2}$ $\mu$m$^{-1}$) received at Earth is calculated by scaling the PHOENIX stellar model to the star's TESS-band magnitude.  We then convert vacuum wavelengths to wavelengths in air as prescribed by \cite{1991ApJS...77..119M}.  For each band, MuSCAT throughput\footnote{\url{http://esppro.mtk.nao.ac.jp/MuSCAT/TM_MuSCAT.dat}}--which includes reflectances and transmittances of the dichroic mirrors, broadband anti-reflection coating on the CCD windows, filters, and quantum efficiencies of the CCDs--is provided in increments of 10 nanometers.  MuSCAT2 throughput is  provided in increments of 0.5 nanometers.  Before applying the throughput to the PHOENIX stellar model, we smooth the stellar model with a Gaussian to produce a spectrum with resolution matching that of the throughput.  As we split the PHOENIX model light into the three bandpasses of MuSCAT, or the four bandpasses of MuSCAT2, we also apply factors for atmospheric transmittance, as well as throughput of mirrors M1 and M2 on the telescopes, as reported by \cite{2015JATIS...1d5001N, 2019JATIS...5a5001N}.  Next, we multiply by the telescope area to produce an output flux across each bandpass in ergs sec$^{-1}$ $\mu$m$^{-1}$.  Finally, we convert this to a photon flux across each bandpass (photons sec$^{-1}$) by dividing by the energy per photon $h\nu$, where $h$ is Planck's constant and $\nu$ is the frequency of the photon, and then numerically integrating with respect to wavelength across each bandpass.

We use the right ascension and declination of the target star to determine whether the system can be observed using MuSCAT or MuSCAT2.  The OAC 188-cm telescope is located at a latitude of 34$^{\circ}$ 34' 37.47" North.  MuSCAT is unable to observe at declinations greater than 75 degrees.  We calculate the fraction of TESS discoveries visible from OAC by assuming that we wish to view through an airmass of 2 or less, which equates to 60 degrees from zenith.  For simplicity, we assume that we view all targeted systems as they cross the meridian, so that only the declination angle enters into our calculation.  Thus, TESS discoveries with declination angles between -25.42 degrees South and +75 degrees North should be visible during some portion of the calendar year.   
    
We perform a similar calculation for MuSCAT2.  The Carlos Sanchez Telescope is located at latitude 28$^{\circ}$ 18' 01.8" North and has physical limits of +64.55$^{\circ}$ North and -36$^{\circ}$ South.  Assuming an airmass limit of 2 and that all targeted systems are observed as they cross the meridian, we determine that TESS discoveries with declination angles between -31.70 degrees South and +64.55 degrees North should be visible using MuSCAT2.  

Noise associated with our observations originates from multiple sources.  In our simulations, we model photon noise from the target star, BEB component stars (false positives only), and sky background, as well as scintillation noise, read noise, and comparison star noise.  

We calculate photon noise for the target star and BEB component stars using

\begin{equation} \label{eq:noise_tgtstar}
N_{\rm star} = \sqrt[]{F_{\rm star}t_{\rm exp}}, 
\end{equation}
\smallskip

\noindent where $F_{\rm star}$ is the photon flux received in a given bandpass from the star, and $t_{\rm exp}$ is the exposure time in seconds.    

We estimate MuSCAT sky background by using the noise values reported for a moonless night by \cite{2015JATIS...1d5001N} during MuSCAT first light observations.  Specifically, we use 19.9 mag arcsec$^{-2}$, 19.5 mag arcsec$^{-2}$, and 18.7 mag arcsec$^{-2}$ for the $g_{2}^{'}$, $r_{2}^{'}$, and $z_{s2}^{'}$ bands, respectively.  We then apply the MuSCAT pixel scale, a Sloan filter conversion tool,\footnote{\url{https://www.gemini.edu/sciops/instruments/midir-resources/imaging-calibrations/fluxmagnitude-conversion}} and the bandpass effective wavelengths and widths reported by \cite{2005ARA&A..43..293B} to convert mag arcsec$^{-2}$ in the MuSCAT bandpasses to an electron noise count per square pixel, $n_{\rm sky}$, which varies with the exposure time duration.  Sky background noise for a given exposure depends upon the photometric aperture on the CCD over which the photons are spread.  We calculate sky background noise for an exposure using

\begin{equation} \label{eq:skynoise}
N_{\rm sky} = \sqrt[]{n_{\rm sky}\pi r_{\rm aper}^2}, 
\end{equation}

\noindent where $r_{\rm aper}$ is the radius of the photometric aperture in pixels.

We calculate sky background for MuSCAT2 using the same methods, but for MuSCAT2 we use the noise values reported for a moonless night by \cite{2019JATIS...5a5001N}, which are 20.4 mag arcsec$^{-2}$, 19.8 mag arcsec$^{-2}$, 19.0 mag arcsec$^{-2}$, and 18.2 mag arcsec$^{-2}$ for $g$, $r$, $i$, and $z_{s}$ bands, respectively.

We calculate scintillation noise using the method described by \cite{1967AJ.....72..747Y} and by \cite{1998PASP..110.1118D}, which is given by

\begin{equation} \label{eq:scin}
N_{\rm scin} = 0.064D^{-2/3}(sec Z)^{7/4}e^{-h/h_o}t_{\rm exp}^{-1/2}F_{\rm star}t_{\rm exp},
\end{equation}
\smallskip

\noindent where $D$ is the diameter of the primary telescope mirror in cm, $Z$ is the local zenith angle of the target star, $h$ is the elevation of the telescope above sea level (372 meters at Okayama Astro-Complex, Japan, and 2,387 meters at Teide Observatory in the Canary Islands), and $h_o = 8000$ meters is a constant.  We use the zenith angle calculated as the target star crosses the meridian.

Read noise varies with MuSCAT bandpass and also depends upon whether fast readout time (0.58 sec) or slow readout time (10 sec) is selected.\footnote{\url{http://esppro.mtk.nao.ac.jp/MuSCAT/observing.html}}  For this study, we assume all observations are conducted using a fast readout time, which results in MuSCAT read noise per square pixel $n_{\rm read}$ of 11, 12, and 12 electrons for the $g_{2}^{'}$, $r_{2}^{'}$, and $z_{s2}^{'}$ bands, respectively.  Similarly, MuSCAT2 read noise per square pixel for fast readout times are 12.35, 11.51, 13.13, and 12.56 electrons for the $g$, $r$, $i$, and $z_{s}$ bands, respectively.  Like sky background noise, read noise for a given exposure depends upon the photometric aperture.  We calculate read noise for an exposure using

\begin{equation} \label{eq:readnoise}
N_{\rm read} = \sqrt[]{n_{\rm read}\pi r_{\rm aper}^2}.
\end{equation}

Observational experience has shown that bright comparison stars are not always available within the MuSCAT/MuSCAT2 fields of view,\footnote{The MuSCAT field of view is 6.1 x 6.1 arcmin,$^2$ while that of MuSCAT2 is 7.4 x 7.4 arcmin.$^2$} so that comparison stars are a non-negligible noise source.\footnote{We analyzed the sensitivity of our results to comparison star noise by calculating results both with and without this noise source.  For MuSCAT (MuSCAT2 values in parentheses), we estimate that including comparison star noise decreases the number of candidate exoplanets that can be distinguished from BEB false positives by $\sim$3\% ($\sim$2\%) for all candidate exoplanets, and by $\sim$2\% ($\sim$1\%) for $R_{\rm pl} < 4R_{\oplus}$ candidate exoplanets. }   However, the actual comparison stars that will be available for any given candidate exoplanet observation are difficult to predict in advance. 

We incorporate comparison star noise into our results by scaling comparison star noise  ($N_{\rm comp}$) recorded during past observations of WASP-12 \citep{2015JATIS...1d5001N,2019JATIS...5a5001N} to the \cite{2018ApJS..239....2B} and TOI systems.  Although our computations using WASP-12 are likely to vary from the true comparison star noise for any given stellar system, we note that our analysis applied to all systems as a whole provides a reasonable estimate of the effects of comparison star noise on our results. This is because the V-band magnitude of WASP-12 is 11.57, and the median V-band magnitude of all \cite{2018ApJS..239....2B} TESS predicted exoplanets is 11.69. Thus, the brightness of WASP-12 roughly corresponds to the median brightness of the \cite{2018ApJS..239....2B} exoplanet systems. For stars that are dimmer than WASP-12, we will likely find more bright comparison stars available for relative photometry, such that comparison star noise will have a lesser effect on the results. However, for stars that are brighter than WASP-12, we will likely find fewer bright comparison stars available, so that comparison star noise will have a greater effect on results.

Finally, we determine the overall effect of these noise sources on our measurement of transit depth in each bandpass.  To do so, we estimate the noise from all sources for a single mid-transit exposure.  We add the stellar photon, sky background, scintillation, and read noise in quadrature, and then divide by total number of photons received from all stars (targeted star and BEB component stars, if applicable) in one exposure to calculate $\sigma_{\rm 1}$, the error associated with the aforementioned noise sources.  We then add $\sigma_{\rm comp}=1/\sqrt{N_{\rm comp}}$ to $\sigma_{\rm 1}$ in quadrature to compute our total estimated error, $\sigma_{\rm total}$.

Our total noise calculation depends upon exposure times.  During actual observations of bright stars, stellar flux is defocused across a larger number of pixels to allow for longer exposure times while still remaining within the linear response regime of the CCD pixels.  Defocusing stellar light also mitigates adverse effects on observations such as scintillation, changing atmospheric conditions, telescope tracking errors and flat-fielding errors \citep{2009MNRAS.396.1023S}.   

Our simulation tool includes an algorithm to optimize telescope defocusing and exposure time for observations of a given system.  We limit exposure times in each band to values between 5 and 60 seconds, and we limit the radius used for telescope defocusing, $r_{\rm aper}$, to values between 3 and 21 pixels for MuSCAT, and to values between 2 and 18 pixels for MuSCAT2.  Our choice of minimum defocusing aperture is motivated by the typical seeing conditions, which are 1.5" at Okayama Astro-Complex, Japan, and 0.8" at Teide Observatory in the Canary Islands.  The maximum aperture radius corresponds to $\sim$15", above which the stellar point spread function becomes asymmetric so that further defocusing is no longer effective.

The defocusing algorithm selects the combination of exposure time and $r_{\rm aper}$ that maintains CCD response in the linear regime\footnote{The MuSCAT linearity range is $<$50,000 ADU for $<$1\% non-linearity, while that for MuSCAT2 is $<$62,000 ADU for $<$1\% non-linearity. } while producing the highest signal-to-noise (S/N) across an entire planetary transit.  We calculate S/N using

\begin{equation} \label{eq:SNRtransit}
S/N_{\rm transit} = {\rm Transit \ Depth} \  \times \ \frac{\sqrt[]{m_{\rm exposure}}}{\sigma_{\rm 1}}.
\end{equation}

\noindent Here, $m_{\rm exposure}$ is the number of exposures recorded during the transit, calculated by dividing the transit duration by the exposure cadence. The exposure cadence is equal to the exposure time plus the dead time per exposure, which accounts for the time required to save data into a proper FITS format and add header information to the FITS file.  Based upon observational experience, we assume a constant 4 sec dead time for each exposure.

\subsection{Comparison of Simulation Results to Observations} \label{sec:validation}

We compared our output MuSCAT/MuSCAT2 simulation results to actual data from \cite{2016ApJ...819...27F}'s MuSCAT observations of HAT-P-14b, as well as \cite{2019JATIS...5a5001N}'s MuSCAT2 observations of WASP-12b.  Initially, our calculations for both target star photons collected and for total noise exceeded those measured during actual observations.  However, we modified our simulation routines to incorporate random factors to account for the largest sources of the differences between our calculated results and observed results, which we deemed to be due to fluctuations in atmospheric transmittance and degradation of mirror reflectance.  We note that our initial simulation output results would lead to better photometric precision, which in turn would bias our results to indicate better ability to distinguish BEB false positives than what would likely be encountered during actual observations.  Thus, application of the random factors is designed to maintain the conservative nature of our results.  In this section, we describe the development of the uniform distributions from which we draw our random factors, and how the random factors are applied in our code. 

For each bandpass of each instrument (MuSCAT or MuSCAT2), we derive a uniform distribution between some minimum value and 1.0 to account for both variations in atmospheric transmittance and degradation of mirror reflectivity.  A value of 1.0 on these distributions represents photometric sky conditions soon after mirror recoating and maintenance.  Conversely, a value near the minimum on these distributions represents poor atmospheric transmittance, with mirror reflectivity degraded by the maximum amount that we consider. 

During a long-term monitoring campaign, \cite{2019AJ....158..206F} recorded the night-to-night variations in relative transmittance of the sky for all bands of both MuSCAT and MuSCAT2.  We use the recorded values as typical variations in atmospheric transmittance for each site.  The data recorded by \cite{2019AJ....158..206F} show correlation between bandpasses.  For example, if atmospheric transmittance was near 1.0, then that was true of all bandpasses.  On the other hand, if the transmittance was near the minimum or near some mid-range value, then that was true of all bandpasses as well. 

Telescope mirror reflectance has been shown to degrade due to chemical reactions and physical effects \citep[e.g.,][]{2016SPIE.9906E..3UA, 2018SPIE10700E..0IH, 2019PASJ...71...32O}.  Mirror recoating and cleaning can offset this degradation, but overall reflectivity has been shown to vary by over 10 percent in the course of a year \citep{2016SPIE.9906E..3UA}.  In our simulations, we assume that reflectivity will degrade by as much as 15\%. 

We create each uniform distribution by taking the variations in atmospheric transmittance found by \cite{2019AJ....158..206F}, and decreasing the minimum values by 0.15 to account for reflectivity degradation.  For example, \cite{2019AJ....158..206F} found that MuSCAT r-band atmospheric transmittance varies between 0.65 to 1.0.  Decreasing the minimum value on this distribution by 0.15 to account for mirror degradation, we use a uniform distribution from 0.50 to 1.0 for MuSCAT r-band.  Similarly, \cite{2019AJ....158..206F} found that MuSCAT2 r-band atmospheric transmittance varies between 0.76 and 1.0.  Decreasing the minimum value on this distribution by 0.15 to account for mirror degradation,  we use a uniform distribution from 0.61 to 1.0 for MuSCAT2 r-band. 

For each trial of our simulation for a given instrument (MuSCAT or MuSCAT2), we draw a random number on a uniform distribution from 0 to 1, $n_{\rm r,0-1}$.  We draw a unique random number for each \cite{2018ApJS..239....2B} or TOI system.  Then, for each bandpass of the instrument, we calculate a random factor using

\begin{equation} \label{eq:randomfactor}
f_{\rm r,band} = v_{\rm min,band} + n_{\rm r,0-1} \times (v_{\rm max,band} - v_{\rm min,band}),
\end{equation}

\noindent where $f_{\rm r,band}$ is the random factor derived for a given bandpass, and $v_{\rm min,band}$ and $v_{\rm max,band}$ are the minimum and maximum values of the uniform distributions created for the corresponding bandpass.  Equation \ref{eq:randomfactor} allows us to model the observed correlation across bandpasses.  For each bandpass and for each stellar system, we multiply the corresponding random factor, $f_{\rm r,band}$, by the number of stellar photons calculated to be collected by the MuSCAT or MuSCAT2 instrument in that bandpass.  

We can apply \textit{ad hoc} factors\footnote{We call these factors \textit{ad hoc} because they are chosen by design, rather than randomly.  However, we select the \textit{ad hoc} factors from the same uniform distributions developed for our random factors. } to ensure that our calculations for target star photons collected at the telescope exactly match those of the \cite{2016ApJ...819...27F} MuSCAT observations of HAT-P-14b and the \cite{2019JATIS...5a5001N} MuSCAT2 observations of WASP-12b.  The \textit{ad hoc} factors that we select for each instrument all lie within the random uniform distributions developed for the bandpasses of MuSCAT and MuSCAT2.  By applying these \textit{ad hoc} factors, our noise calculations for HAT-P-14b also match those of \cite{2016ApJ...819...27F} to within 3\% for every band.  For MuSCAT2, our noise calculations match those of \cite{2019JATIS...5a5001N} to 7\%, 6\%, 3\%, and $<$1\% for the $g$, $r$, $i$, and $z_{s}$ bands.  This shows that application of random factors drawn from the uniform distributions that we developed are more likely to produce realistic results. 

\subsection{Blended Eclipsing Binary False Positives} \label{sec:FP}

We construct false positives for all simulated TESS discoveries in the \cite{2018ApJS..239....2B} machine-readable file such that the transit depths for both the false positive and the transiting exoplanet are the same in the TESS bandpass.  During the actual TESS mission, some \textit{candidate} exoplanets are true exoplanets, while some are false positives that mimic the signal of a transiting exoplanet.  Here, we use the \cite{2018ApJS..239....2B} simulated TESS discoveries as a representative sample of the types of signals that TESS may discover.  After creating the false positives, we determine whether MuSCAT and MuSCAT2 can distinguish varying transit depths between bandpasses, as described in Section \ref{sec:MuSCATcriterion}.   

To create BEB false positives, we begin by calculating the transit depth for a true exoplanet using the relationship

\begin{equation} \label{eq:trandepthEXOPL}
\frac{\Delta F}{F} = \frac{R_{\rm pl}^2}{R_{\rm star}^2},
\end{equation}

\noindent where $\Delta F$ refers to the difference between out-of-transit and mid-transit flux, while $R_{\rm pl}$ and $R_{\rm star}$ refer to the radii of the planet and of the target star, respectively.  By calculating the exoplanet transit depth using equation \ref{eq:trandepthEXOPL}, we inherently assume that the planet emits no flux of its own.

We pick the primary and secondary components of the BEBs from data files of simulated stars downloaded from the TRIdimensional modeL of thE GALaxy, or TRILEGAL  \citep{2012ASSP...26..165G}.  We downloaded selections of TRILEGAL stars\footnote{\url{http://stev.oapd.inaf.it/cgi-bin/trilegal}} at Galactic coordinates with Longitudes of 0, 30, 60, 90, 120, 150, and 180 degrees, and at Latitudes of 0, 30, 60, and 90 degrees.  We chose solid angles ranging from 0.0001 deg$^2$ to 1 deg$^2$, depending upon proximity to the Galactic center.  In every case, the data files we downloaded for each Galactic coordinate pair contained selections of over 10,000 stars.  We assume symmetry in the Galactic coordinates, and then choose BEB components from the TRILEGAL data file corresponding to Galactic coordinates closest to those of the target star under consideration.  The TRILEGAL data files include important parameters for the stars, such as masses, log(g), effective temperatures, distances, and TESS-band apparent magnitudes.

We choose the primary BEB component such that the star lies on the main sequence, and so that its apparent magnitude is greater than that of the target star apparent magnitude (i.e. the primary BEB component is fainter than the target star).  We randomly pick the primary BEB component from the appropriate TRILEGAL data file, making our selection from stars with maximum and minimum TESS-band apparent magnitudes between the values of 

\begin{equation} \label{eq:magBEBprimax}
m_{\rm BEB,pri,max} = m_{\rm star} - 2.5\ {\rm log}\ \Bigg(\frac{R_{\rm pl}^2}{R_{\rm star}^2}\Bigg),
\end{equation}

\noindent and

\begin{equation} \label{eq:magBEBprimin}
m_{\rm BEB,pri,min} = m_{\rm star} - 0.5\ {\rm log}\ \Bigg(\frac{R_{\rm pl}^2}{R_{\rm star}^2}\Bigg),
\end{equation}

\noindent where $m_{\rm BEB,pri,max}$ and $m_{\rm BEB,pri,min}$ are our maximum and minimum TESS-band apparent magnitude limits for the BEB primary component, and $m_{\rm star}$ is the target star TESS-band apparent magnitude.  Here, the maximum apparent magnitude limit is chosen such that the ratio of the primary component's flux to that of the target star is equal to the transit depth calculated using equation \ref{eq:trandepthEXOPL}.  We could expect a primary component magnitude near this value if the primary component star is totally eclipsed during transit (i.e. if primary and secondary components are the same size).  Our apparent magnitude selection limits are designed to give us a wide selection of choices for BEB primary components from the TRILEGAL file.  The number of choices actually available will depend upon the transit depth in the system under consideration.  For most systems, the magnitude cuts provide hundreds of choices for primary BEB component star with masses $\sim$1M$_\odot$ or less.  Even the systems with the largest transit depths ($\sim$0.1) provide $\sim$100 choices for the BEB primary component star. 

Following the method of \cite{2019AJ....157...77G}, we next select the \textit{desired} mass of the secondary component by randomly selecting a mass on a uniform distribution between 0.1$M_{\odot}$ to the mass of the primary BEB component.  We then pick a star from the TRILEGAL catalog that matches this desired secondary component mass.  We note that other authors have used different distributions for the secondary component masses \citep [e.g.,][]{2016ApJ...822...86M}.  However, the precise mass function for BEB components is uncertain, and the uniform distribution used here provides a reasonable estimate that is suitable for our purposes.

After the primary and secondary components to the BEB are selected, we  calculate the transit depth caused by the BEB system using

\begin{equation} \label{eq:trandepthBEB}
\frac{\Delta F}{F} = \frac{F_{\rm OOT} - F_{\rm MT}}{F_{\rm OOT}},
\end{equation}

\noindent where $F_{\rm OOT}$ is the out-of-transit flux and $F_{\rm MT}$ is the mid-transit flux.  We calculate out-of-transit flux by summing the total flux from the target star and two BEB component stars.  We calculate mid-transit flux by summing total flux from the target star and BEB secondary star, but we decrease the amount of flux from the BEB primary component star by the factor $R_{\rm BEB,sec}^2/R_{\rm BEB,pri}^2$.  Note that our method assumes all BEBs are observed edge-on.  We adjust the distance to the BEB such that the transit depths calculated using equations \ref{eq:trandepthEXOPL} and \ref{eq:trandepthBEB} (in the TESS bandpass) match to within machine accuracy.  The adjusted distance is close to that of the original TRILEGAL distance, thus preserving the integrity of the TRILEGAL sample.  We then calculate the TESS-band magnitudes of the primary and secondary BEB components corresponding to the newly calculated distance of the BEB, and apply our simulation tools as described in section \ref{sec:simtool} to determine the noise produced in the MuSCAT or MuSCAT2 bandpasses for each BEB component star, in addition to that produced by the target star.  We compute transit depths in all MuSCAT and MuSCAT2 bandpasses by applying equation \ref{eq:trandepthBEB} to our estimated flux (Section \ref{sec:simtool}) for each bandpass.

\subsection{Distinguishing BEBs with MuSCAT/MuSCAT2} \label{sec:MuSCATcriterion}

We use a simple computational model to determine whether MuSCAT/MuSCAT2 can discriminate between a true exoplanet and a BEB false positive.  We expect the transit depth for a BEB false positive to vary approximately linearly with wavelength, such that the transit depths in MuSCAT's 3 bandpasses--or in MuSCAT2's 4 bandpasses--can be reasonably fit with a straight line.  The MuSCAT/MuSCAT2 simulation tools (Section \ref{sec:simtool}) applied to BEBs (Section \ref{sec:FP}) produce estimated transit depths and noise values, $\sigma_{\rm total}$, for each MuSCAT bandpass.  For each system, we use the estimated transit depths and noise values to find the best-fit linear least squares line, and then we determine whether that line can be distinguished from a flat line.  We would expect the best-fit line to be flat for a true exoplanet if the presence of a planetary atmosphere is neglected.  Thus, a non-zero slope in the best-fit line should indicate that the transit depths are those of a BEB false positive.

Mathematically, we find that MuSCAT or MuSCAT2 can discriminate a BEB if

\begin{equation} \label{eq:distinguishBEB}
\mid {\rm slope}\mid - \ 3 \ \times \mid {\rm slope_{err}}\mid \ > \ 0,
\end{equation}

\noindent where $slope$ is the slope of the line determined by the least squares fit, and $slope_{err}$ is the error in that slope.  We multiply the slope error by 3 to ensure that the slope of the best-fit line can be clearly distinguished from a flat line.

\section{Results and Discussion} \label{sec:results}

    We applied our MuSCAT and MuSCAT2 simulation tools to the \cite{2018ApJS..239....2B} predicted TESS planet yield to determine the characteristics of those planetary systems where the two instruments can best distinguish between true exoplanets and BEB false positives (Section \ref{sec:MuSCATability}). Next, we applied our simulation tools to TESS Objects of Interest (TOIs) posted to the Exoplanet Follow-up Observing Program for TESS (ExoFOP-TESS) website\footnote{\url{https://exofop.ipac.caltech.edu/tess/} accessed on 16 August 2019} to predict the probabilities that recent TOIs can be distinguished from BEB false positives using MuSCAT and MuSCAT2 (Section \ref{sec:ApplicationToTESSdata}).  As discussed in section \ref{sec:FP}, our simulation randomly generates BEBs each time the routine is applied to a group of planetary systems, and then adjusts the distance to those BEBs so that the transit depth of each BEB false positive matches that of a true transiting exoplanet in the TESS bandpass.  The ability of a given instrument to actually distinguish a false positive for a given system will depend upon the characteristics of the BEB in the system, which is something we will not know \textit{a priori}.  In order to account for natural variation in BEB component characteristics and attain statistically robust results, for both the \cite{2018ApJS..239....2B} planetary systems and the TOIs we report the results attained over 20 trials.  We conducted 20 trials to ensure that our reported results do not vary by more than 1\% throughout our computed 99\% confidence regions \citep{2000ipse.book.....R}.  As described in section \ref{sec:MuSCATcriterion}, the criterion we use to determine whether or not MuSCAT or MuSCAT2 can distinguish a true exoplanet from a false positive relies upon determining the best-fit line to the transit depths.  For MuSCAT, we found the best-fit line using all bandpasses.  For MuSCAT2, we found the best-fit line using the $g$, $r$, and $z_{s}$ bands, since these correspond to the three bandpasses of MuSCAT.  Determining the best-fit line using three bandpasses allows more direct comparison of the MuSCAT and MuSCAT2 results.   

\begin{figure*}
\gridline{\fig{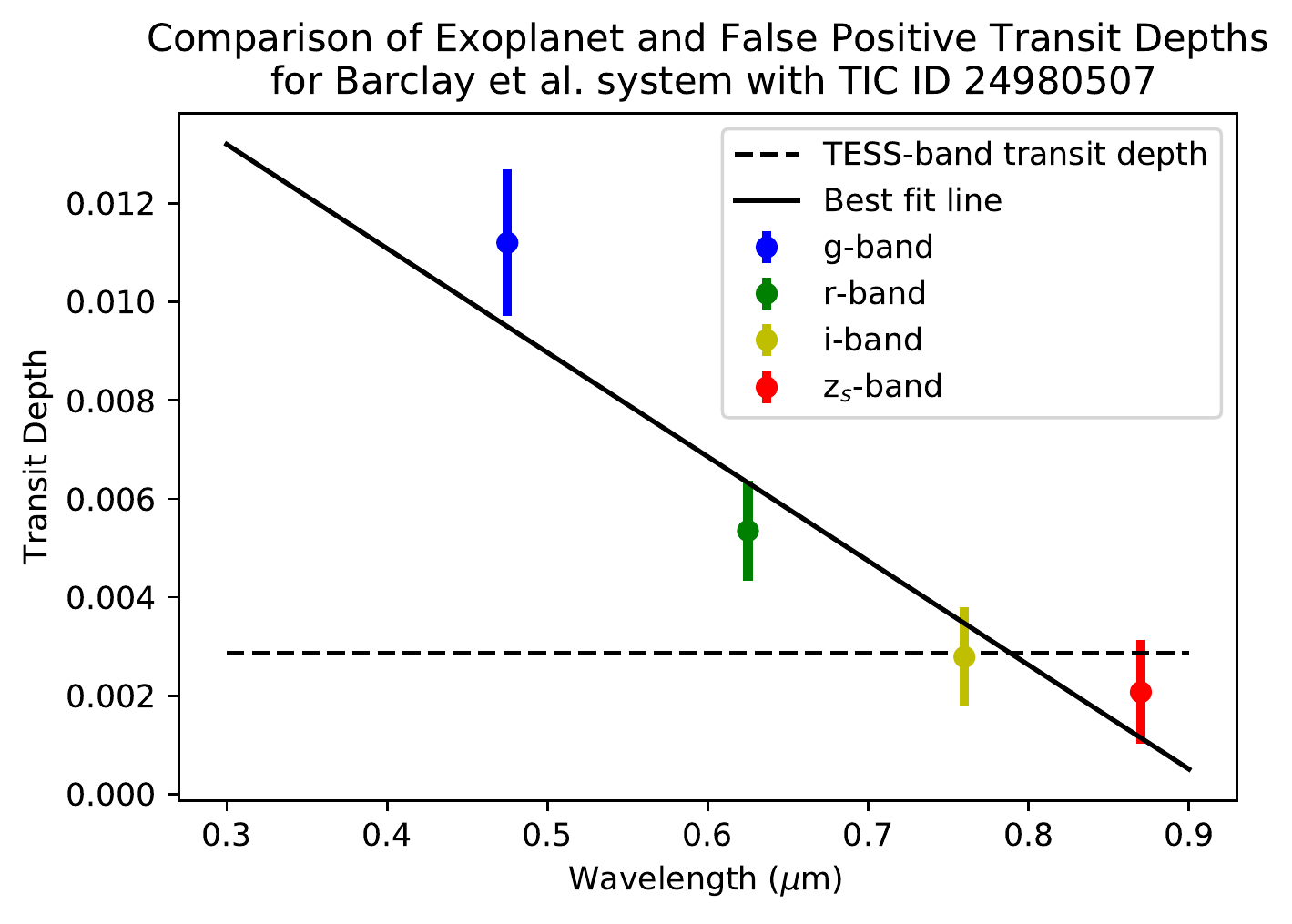}{0.5\textwidth}{(a)}
          \fig{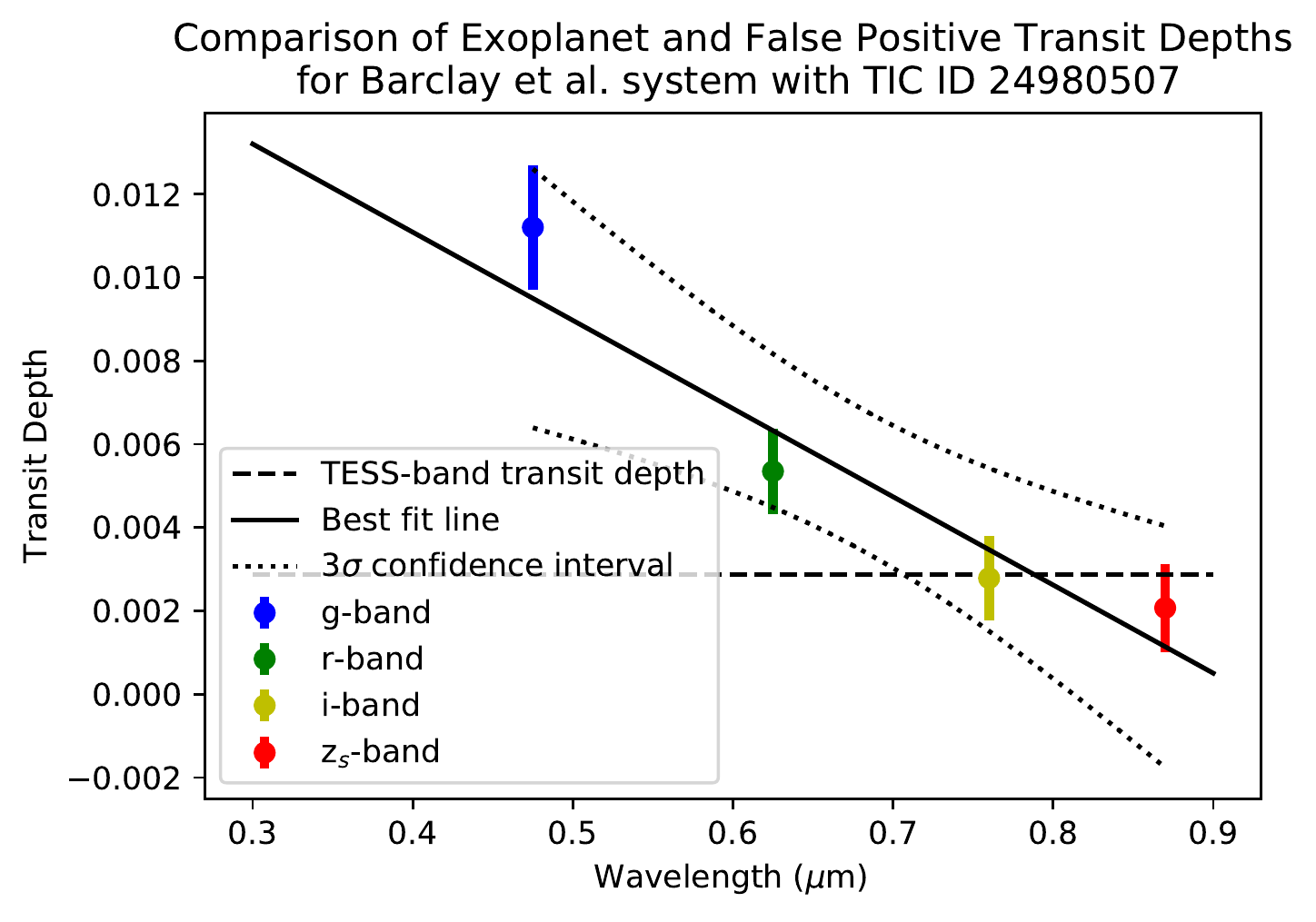}{0.5\textwidth}{(b)}}
\gridline{\fig{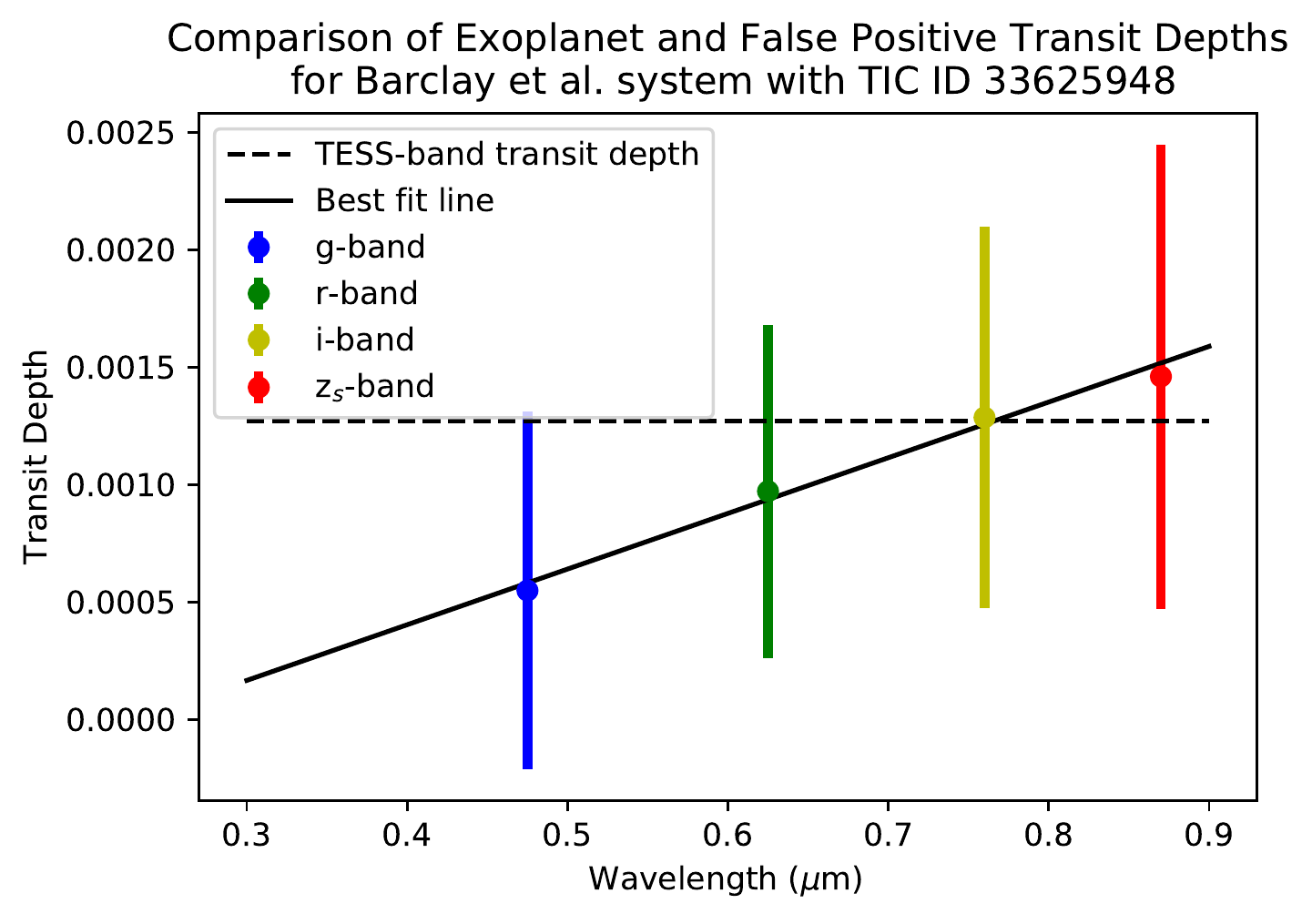}{0.5\textwidth}{(c)}
          \fig{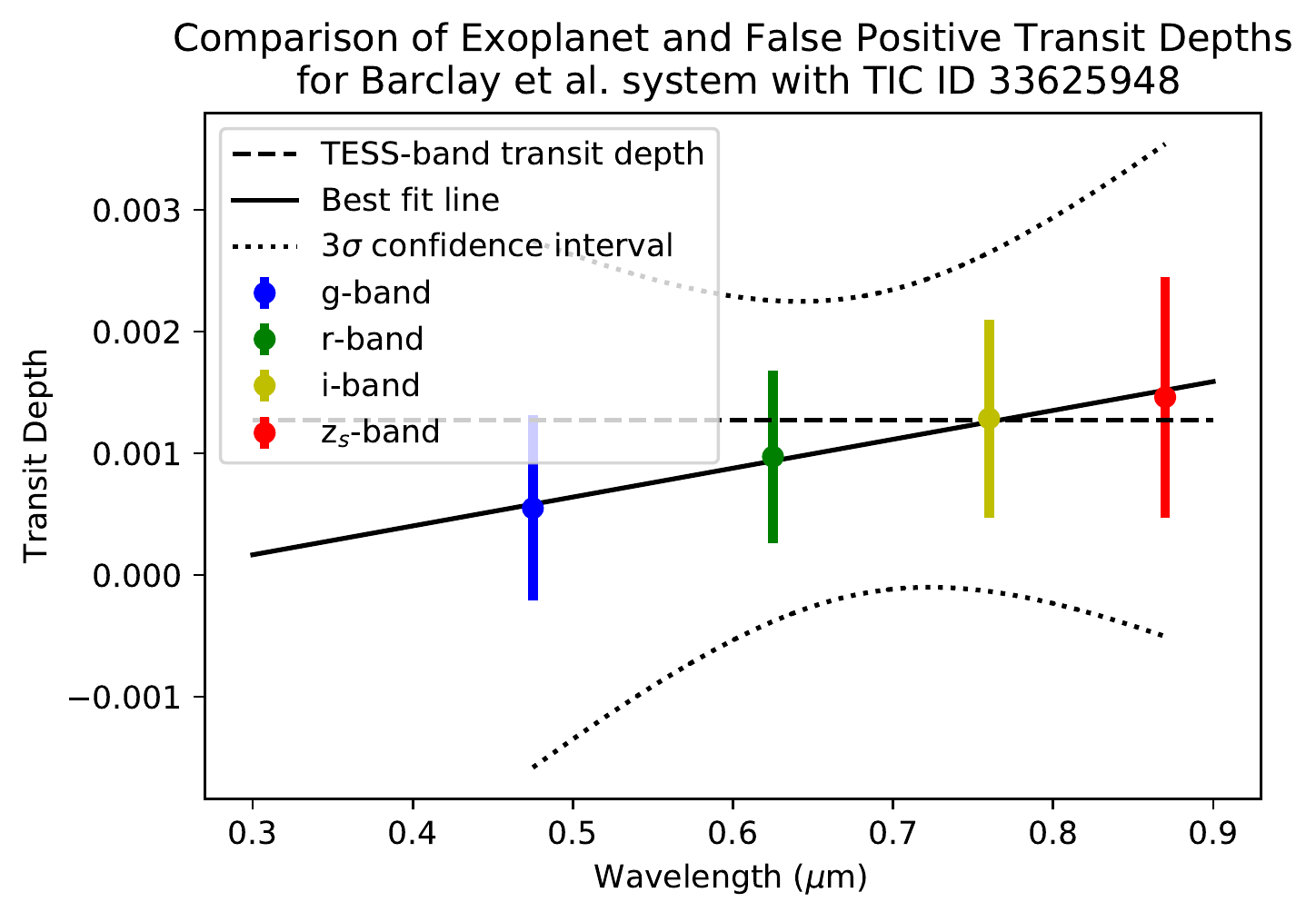}{0.5\textwidth}{(d)}}
\caption{Two examples showing estimated transit depth in MuSCAT2's 4 bandpasses for a BEB, as well as the least squares best-fit line to those transit depths.  For comparison, the TESS-band transit depths for a true exoplanet are also shown.  In panels (b) and (d), we also plot the $3\sigma$ hyperbolic confidence bands \citep{doi:10.1111/j.1751-5823.2007.00027.x} for the best-fit line to help illustrate the criterion we use to determine whether MuSCAT2 can discriminate between a BEB false positive and a true exoplanet.  If MuSCAT2 \textit{cannot} distinguish the BEB, then the line for the TESS-band transit depth falls within the hyperbolic confidence band.  (Note that our criterion in equation \ref{eq:distinguishBEB} is actually much simpler than the hyperbolic confidence bands, which we use here for illustrative purposes only.)  In panels (a) and (b), the magnitude of the slope of the best fit line exceeds the $3\sigma$ slope error magnitude (equation \ref{eq:distinguishBEB}), so we determine that MuSCAT2 \textit{can} discriminate the BEB.  In panels (c) and (d), MuSCAT2 is \textit{unable} to discriminate the BEB.  See text for further details regarding characteristics of these two systems.\label{fig:DistinguishingBEBs}}
\end{figure*}

Figure \ref{fig:DistinguishingBEBs} illustrates the criterion that we use to determine whether MuSCAT and MuSCAT2 can distinguish BEB false positives.  We show two example cases: one where MuSCAT can distinguish the BEB (Figures  \ref{fig:DistinguishingBEBs}a and \ref{fig:DistinguishingBEBs}b), and one where it cannot (Figures \ref{fig:DistinguishingBEBs}c and \ref{fig:DistinguishingBEBs}d).  The system shown in parts (a) and (b) has a transit depth of $\sim$0.00287 in the TESS bandpass, a transit duration of 2.594 hours, and a J-band magnitude of 10.42.  The target star radius is 0.517$R_{\odot}$.  If the transit depth were produced by a true exoplanet, that planet would have a radius of 3.031$R_{\oplus}$.  We created a BEB by adding a primary component star of mass 0.978$M_{\odot}$, and a secondary component star of mass 0.254$M_{\odot}$.  The system in parts (c) and (d) of the figure has a transit depth of $\sim$0.0127 in the TESS bandpass, a transit duration of 4.944 hours, and a J-band magnitude of 10.65.  The target star radius is 1.2$R_{\odot}$.  If the transit depth were produced by a true exoplanet, that planet would have a radius of 4.682$R_{\oplus}$.  We created a BEB by adding a primary component star of mass 0.439$M_{\odot}$, and a secondary component star of mass 0.424$M_{\odot}$.

\subsection{MuSCAT/MuSCAT2 Ability to Distinguish BEBs} \label{sec:MuSCATability}

We report the ability of MuSCAT and MuSCAT2 to distinguish BEBs for the \cite{2018ApJS..239....2B} systems in Tables \ref{tab:MuSCATResults} and \ref{tab:MuSCAT2Results}, respectively.  Overall, we find that MuSCAT is able to discriminate BEBs for $\sim$17\% of the 2,575 \cite{2018ApJS..239....2B} systems observable from the Okayama Astro-Complex (OAC), while MuSCAT2 is able to discriminate BEBs for $\sim$18\% of the 2,485 systems observable from Teide Observatory.  Of the systems visible from OAC, 1,306 have transit depths corresponding to planets with radii less than 4$R_{\oplus}$, and MuSCAT would be able to discriminate $\sim$13\% of these systems.  For MuSCAT2, 1,212 systems have transit depths corresponding to planets with radii less than 4$R_{\oplus}$, and MuSCAT2 would be able to distinguish $\sim$15\% of these systems.  

\begin{deluxetable*}{cccc}
\tablecaption{Predicted number of \cite{2018ApJS..239....2B} systems where MuSCAT can distinguish true exoplanets from BEB false positives\tablenotemark{i} \label{tab:MuSCATResults}}
\tablewidth{0pt}
\tablehead{
\colhead{} & \colhead{Total} & \colhead{Year 1} & \colhead{Year 2} \\
\colhead{} & \colhead{} & \colhead{Ecliptic South} & \colhead{Ecliptic North}}
\startdata
Barclay et al. \\ Candidate Exoplanets & 4,373 & 2,196 & 2,177 \vspace{2mm}\\
Observable from \\ Okayama Astro-Complex\tablenotemark{ii} & 2,575 (59\%) & 539 (25\%) & 2,036 (94\%)  \vspace{2mm} \\
{MuSCAT Can Distinguish\tablenotemark{iii}} & 426 $\pm$ \ 10 & 98 $\pm$ \ 6 & 329 $\pm$ \ 9 \\
{(Total)} & (17\%) & (18\%) & (16\%) \vspace{2mm} \\
{MuSCAT Can Distinguish\tablenotemark{iv}} & 173 $\pm$ \  7  & 39 $\pm$ \ 3 &  134 $\pm$ \ 7 \\
($R_{\rm pl} < 4.0R_{\oplus}$) & (13\%) & (16\%) & (13\%) \vspace{2mm} \\
\enddata
\tablenotetext{i}{\ We report the mean number of planets distinguishable over 20 trials, as well as the 99\% confidence intervals \citep{2000ipse.book.....R}. Each mean is rounded to the nearest whole number of systems, while each confidence interval is rounded up to an integer value.}
\tablenotetext{ii}{\ We assume the system is observable if it is visible through 2 airmasses or less at some point during the calendar year.  See section \ref{sec:simtool} for details.  Percentages in parentheses refer to percent of total Barclay et al. candidate exoplanets in first row.}
\tablenotetext{iii}{\ We report the mean number MuSCAT can distinguish, plus or minus the 99\% confidence intervals over 20 trials.  The number in parentheses is the percentage of the total observable (second row) that MuSCAT can distinguish.}
\tablenotetext{iv}{\ The number in parentheses is the percentage of observable $R_{\rm pl} < 4.0R_{\oplus}$ candidate planets (1,306 total) that MuSCAT can distinguish.  Note that 243 $R_{\rm pl} < 4.0R_{\oplus}$ candidates are observable during year 1, and 1,063 candidates are observable during year 2.}\end{deluxetable*}

\begin{deluxetable*}{cccc}
\tablecaption{Predicted number of \cite{2018ApJS..239....2B} systems where MuSCAT2 can distinguish true exoplanets from BEB false positives\tablenotemark{i} \label{tab:MuSCAT2Results}}
\tablewidth{0pt}
\tablehead{
\colhead{} & \colhead{Total} & \colhead{Year 1} & \colhead{Year 2} \\
\colhead{} & \colhead{} & \colhead{Ecliptic South} & \colhead{Ecliptic North}}
\startdata
Barclay et al. \\ Candidate Exoplanets & 4,373 & 2,196 & 2,177 \vspace{2mm}\\
Observable from \\ Teide Observatory\tablenotemark{ii} & 2,485 (57\%) & 721 (33\%) & 1,764 (81\%)  \vspace{2mm} \\
{MuSCAT2 Can Distinguish\tablenotemark{iii}} & 456 $\pm$ \ 12 & 142 $\pm$ \ 7 & 314 $\pm$ \ 11 \\
{(Total)} & (18\%) & (20\%) & (18\%) \vspace{2mm} \\
{MuSCAT2 Can Distinguish\tablenotemark{iv}} & 180 $\pm$ \  8  & 59 $\pm$ \ 5 &  121 $\pm$ \ 6 \\
{($R_{\rm pl} < 4.0R_{\oplus}$)} & (15\%) & (18\%) & (14\%) \vspace{2mm} \\
\enddata
\tablenotetext{i}{\ We report the mean number of planets distinguishable over 20 trials, as well as the 99\% confidence intervals \citep{2000ipse.book.....R}. Each mean is rounded to the nearest whole number of systems, while each confidence interval is rounded up to an integer value.}
\tablenotetext{ii}{\ We assume the system is observable if it is visible through 2 airmasses or less at some point during the calendar year.  See section \ref{sec:simtool} for details.  Percentages in parentheses refer to percent of total Barclay et al. candidate exoplanets in first row.}
\tablenotetext{iii}{\ We report the mean number MuSCAT2 can distinguish, plus or minus the 99\% confidence intervals over 20 trials.  The number in parentheses is the percentage of the total observable (second row) that MuSCAT2 can distinguish.}
\tablenotetext{iv}{\ The number in parentheses is the percentage of observable $R_{\rm pl} < 4.0R_{\oplus}$ candidate planets (1,212 total) that MuSCAT2 can distinguish.  Note that 321 $R_{\rm pl} < 4.0R_{\oplus}$ candidates are observable during year 1, and 891 candidates are observable during year 2.}
\end{deluxetable*}

\begin{figure*}
\gridline{\fig{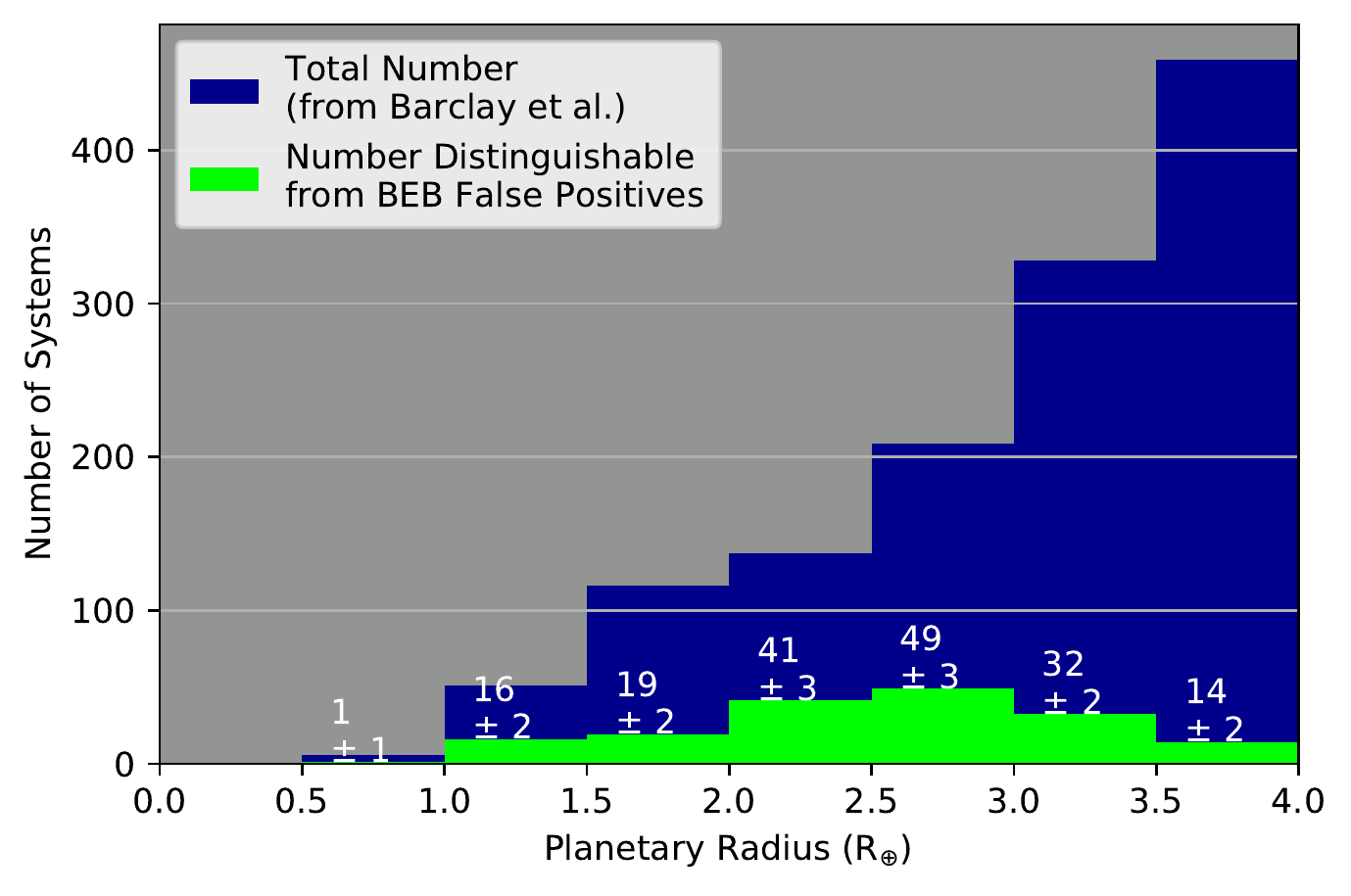}{0.5\textwidth}{(a)}
          \fig{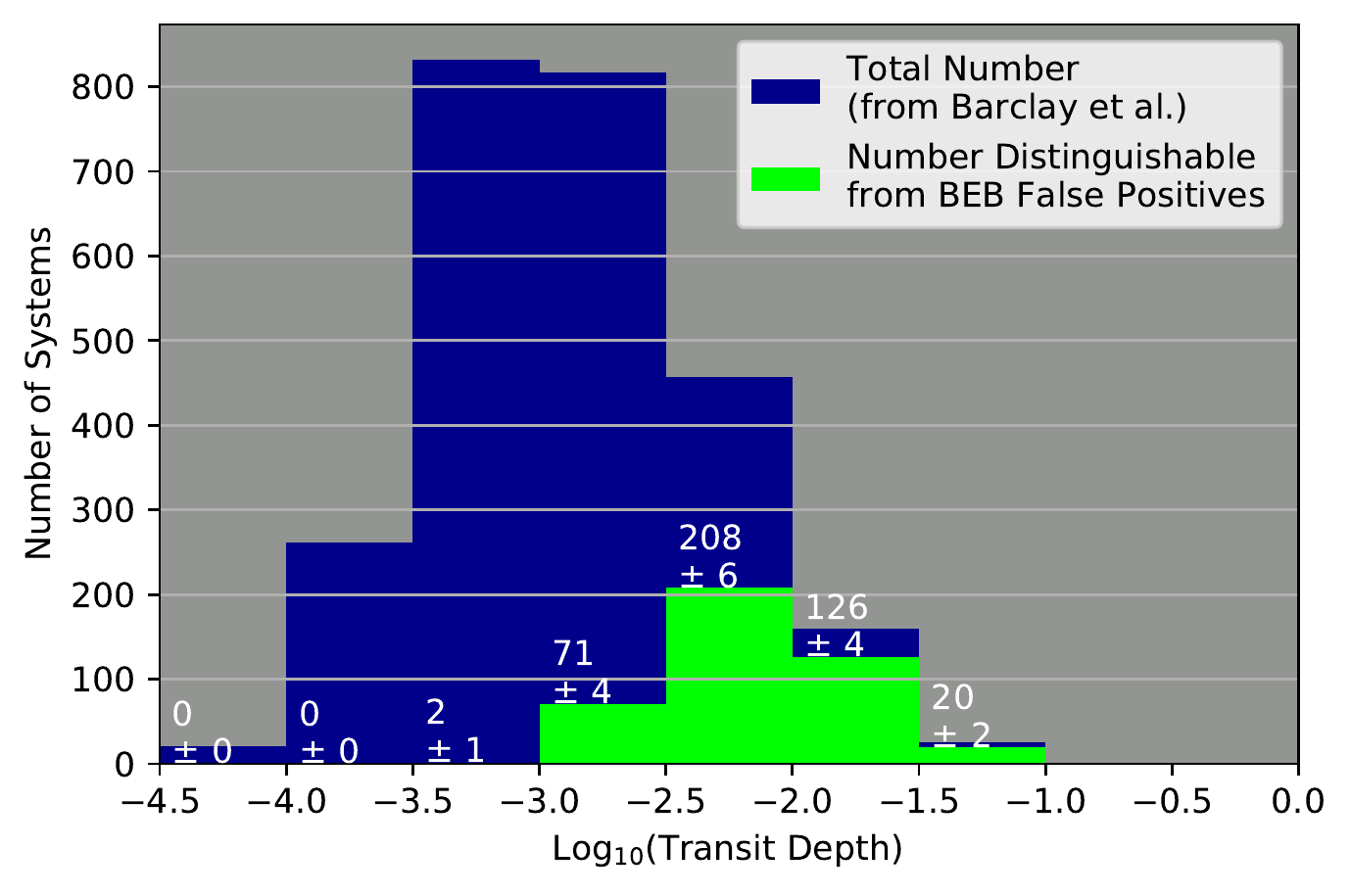}{0.5\textwidth}{(b)}}
\gridline{\fig{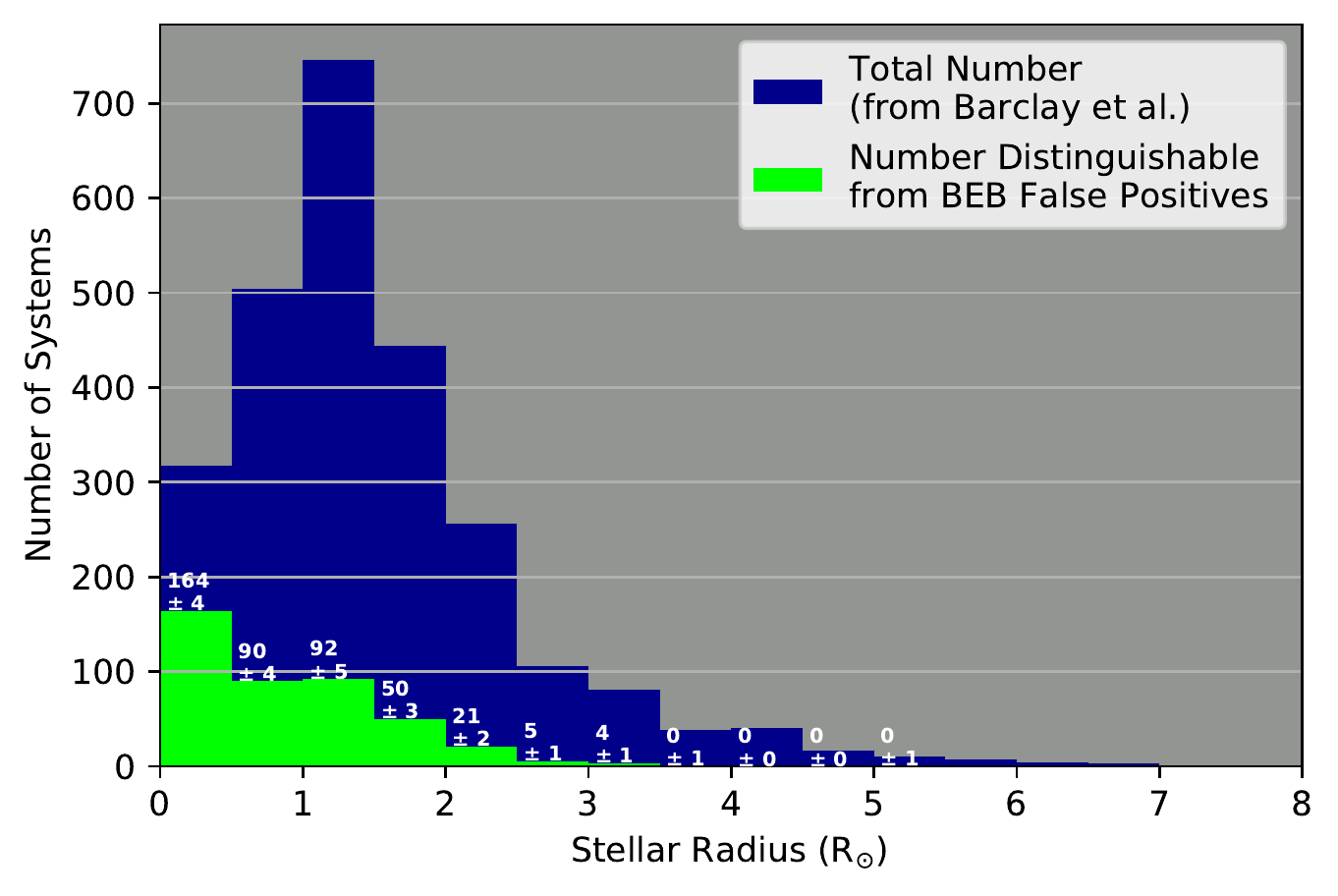}{0.5\textwidth}{(c)}
          \fig{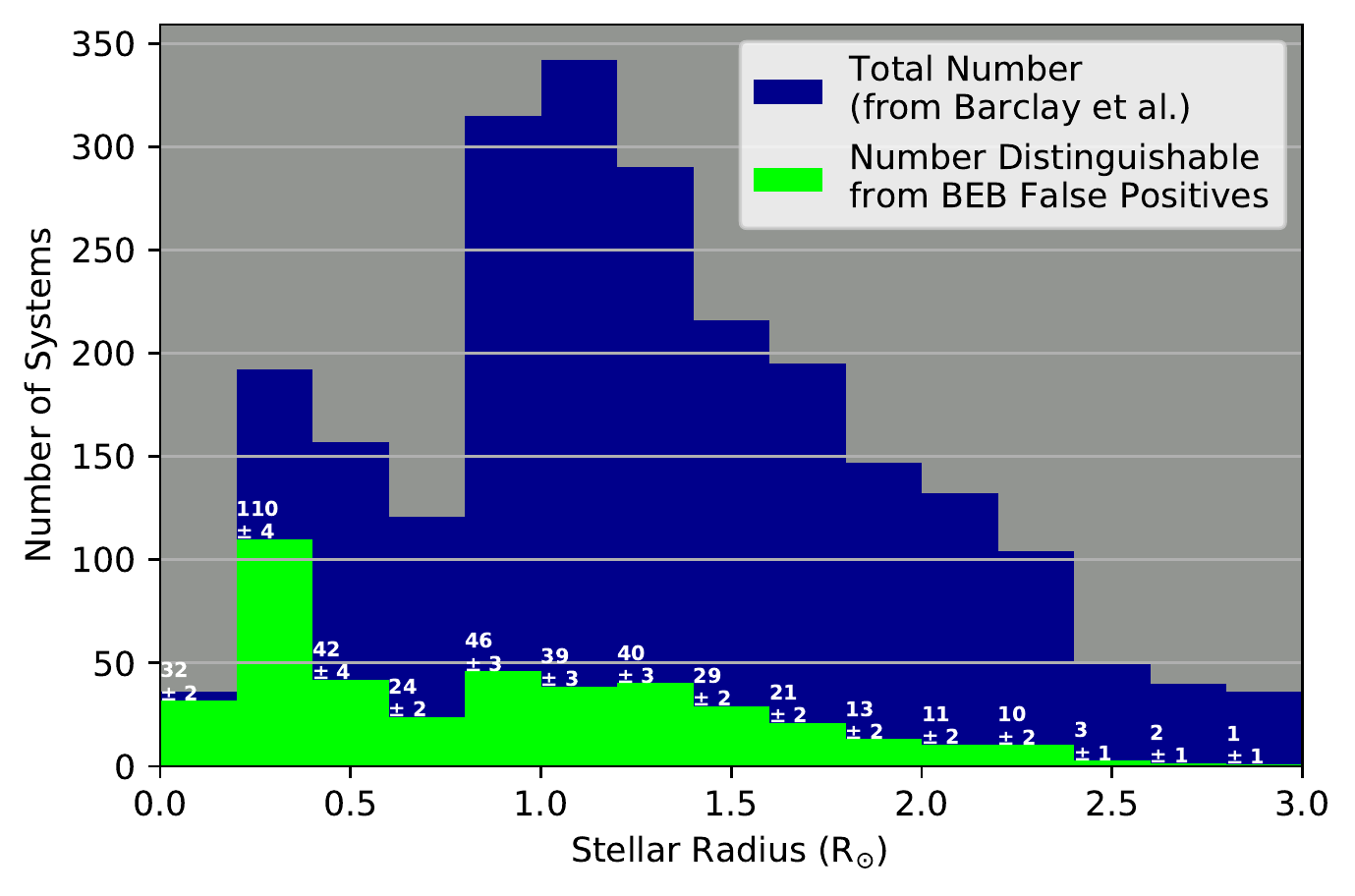}{0.5\textwidth}{(d)}}
\gridline{\fig{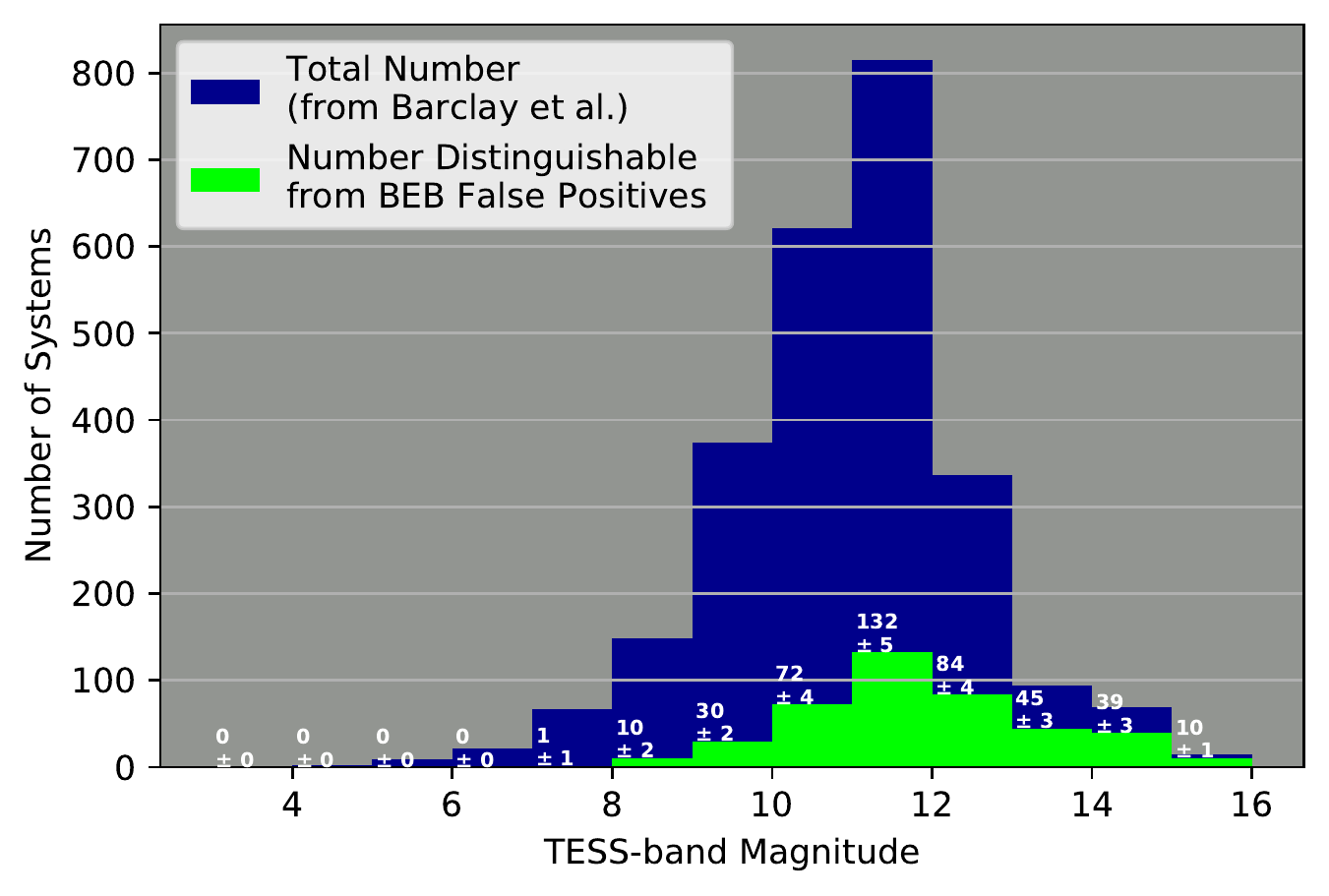}{0.5\textwidth}{(e)}}
\caption{Histograms comparing the total number of \cite{2018ApJS..239....2B} systems visible from Okayama Astro-Complex (2,575 systems) to those that would be distinguishable as BEB false positives using MuSCAT.  We plot the number of systems distinguishable over 20 trials versus (a) planetary radius, (b) transit depth, (c) and (d) stellar radius, and (e) TESS-band magnitude.  We print the average number of distinguishable systems over 20 trials above each bin.  Note that both panels (c) and (d) show the number of systems versus stellar radius, but panel (d) shows this information on a refined grid only for $R_{\rm star} \leq 3R_{\odot}$.  The fraction of systems distinguishable does \textit{not} appear to rise with planetary radius.  However, the fraction of distinguishable systems \textit{does} rise with increasing transit depth and with decreasing stellar radius.  The large fraction of systems distinguishable at higher values of TESS magnitude is due to the fact that transit depths in general are larger for dimmer TESS detections. For example, the median transit depth for systems visible from Okayama Astro-Complex that have TESS magnitude greater than 13 is 0.00879, while the median transit depth for systems brighter than TESS magnitude 13 is 0.00112. \label{fig:MuSCATDistinguishable}}
\end{figure*}

\begin{figure*}
\gridline{\fig{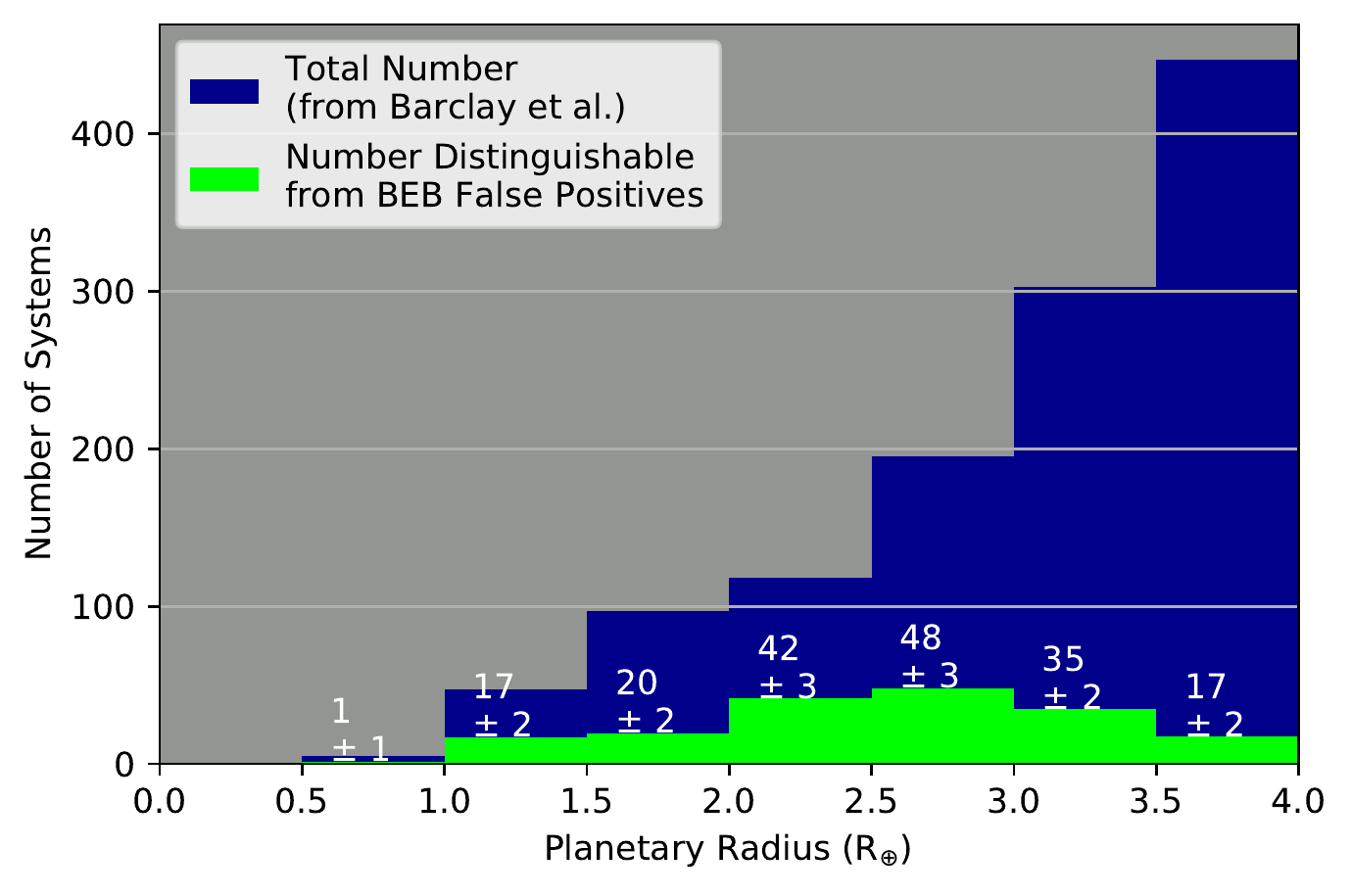}{0.5\textwidth}{(a)}
          \fig{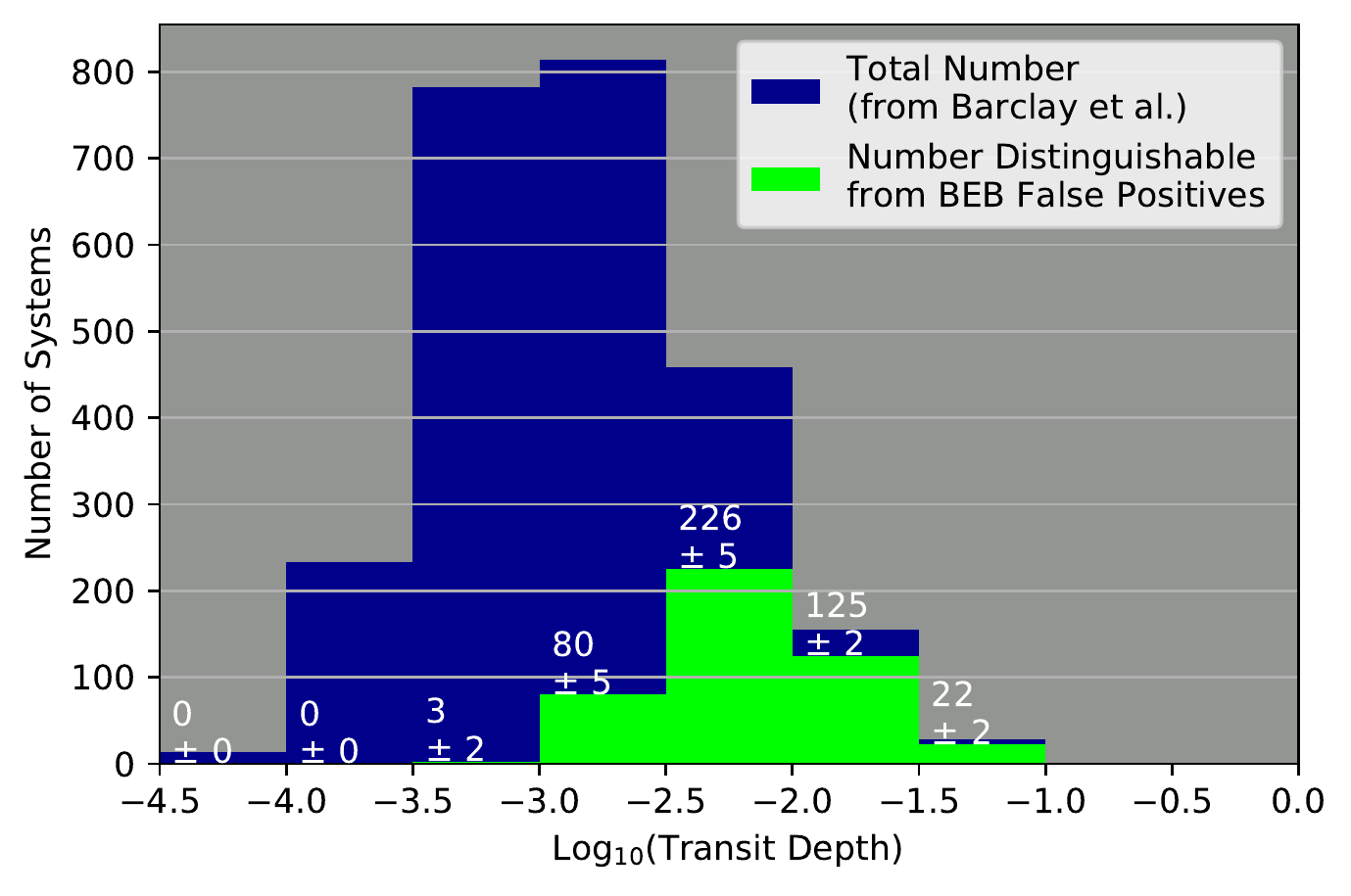}{0.5\textwidth}{(b)}}
\gridline{\fig{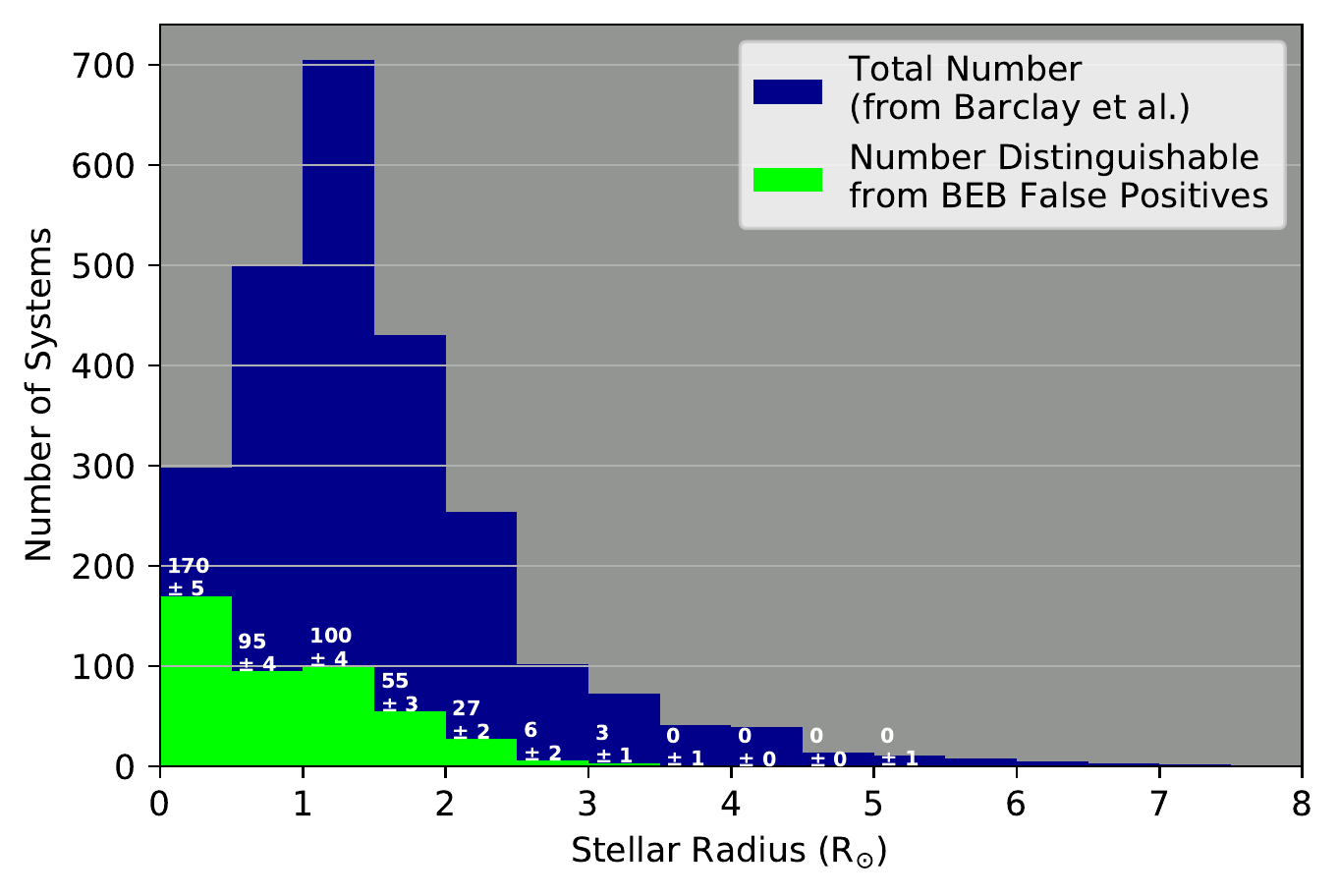}{0.5\textwidth}{(c)}
          \fig{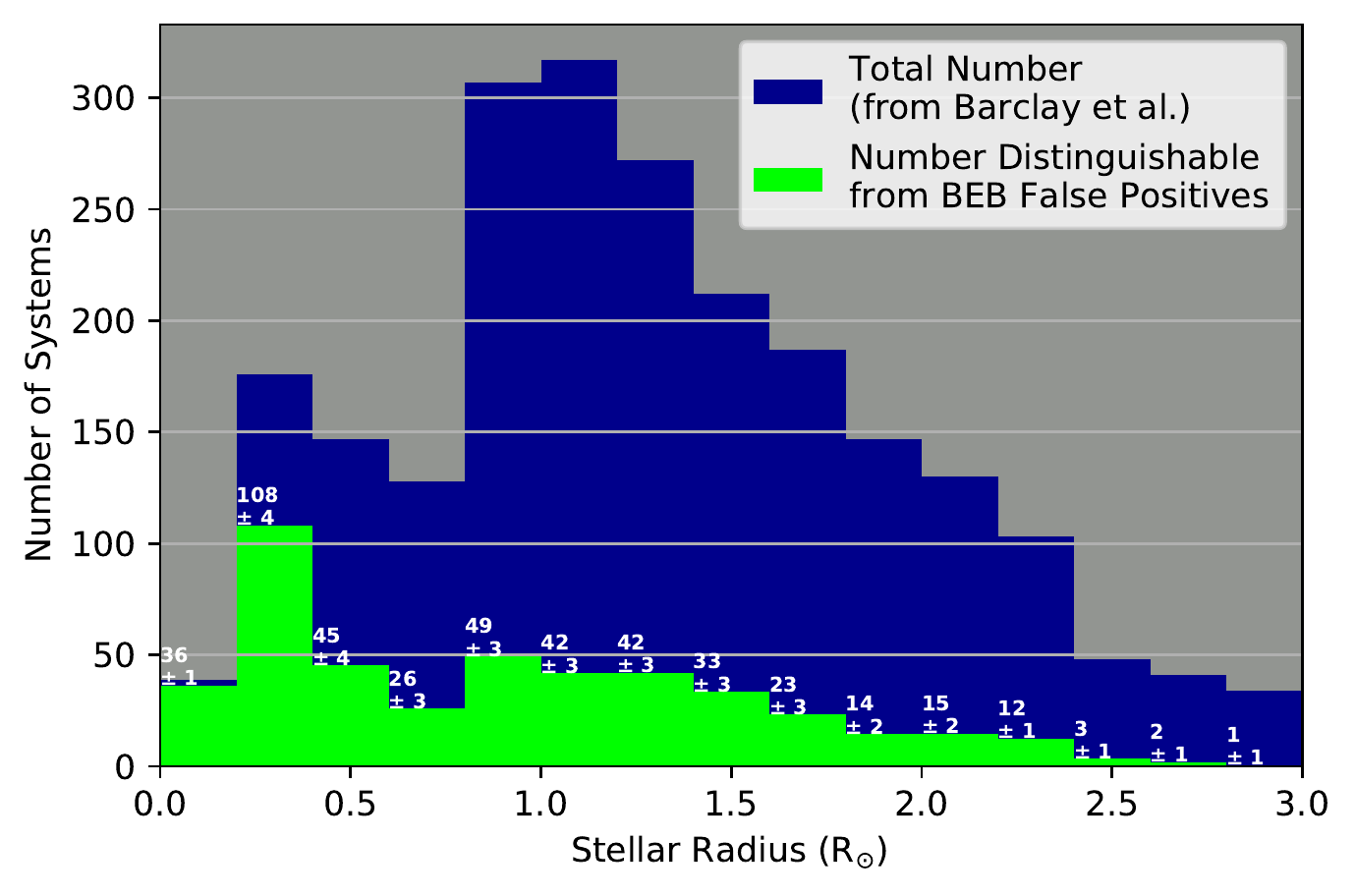}{0.5\textwidth}{(d)}}
\gridline{\fig{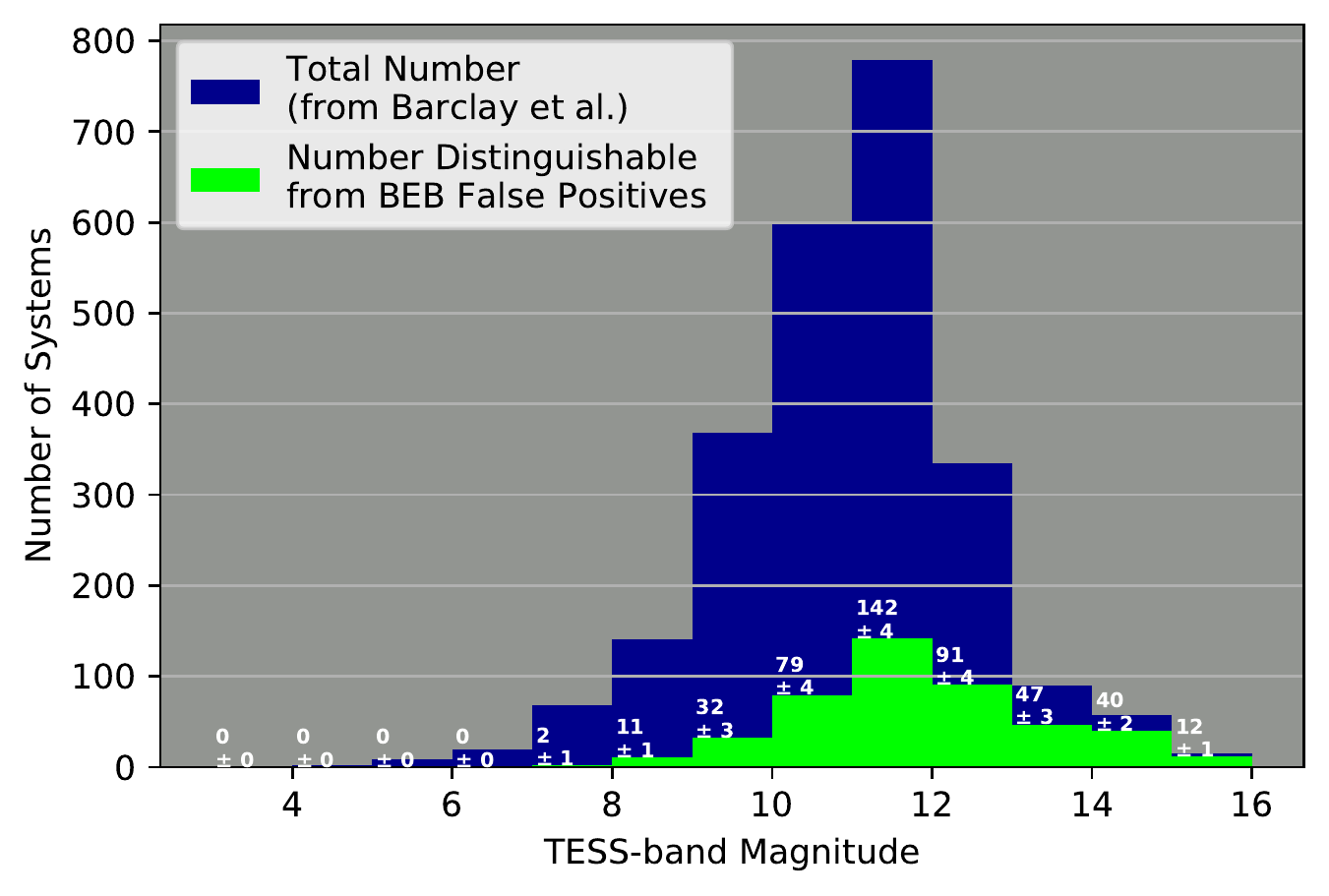}{0.5\textwidth}{(e)}}
\caption{Same as Figure \ref{fig:MuSCATDistinguishable}, but for MuSCAT2.  Histograms comparing the total number of \cite{2018ApJS..239....2B} systems visible from Teide Observatory (2,485 systems) to those that would be distinguishable as BEB false positives using MuSCAT2.  We plot the number of systems distinguishable over 20 trials versus (a) planetary radius, (b) transit depth, (c) and (d) stellar radius, and (e) TESS-band magnitude.  We print the average number of distinguishable systems over 20 trials above each bin.  Note that both panels (c) and (d) show the number of systems versus stellar radius, but panel (d) shows this information on a refined grid only for $R_{\rm star} \leq 3R_{\odot}$.  As with MuSCAT, the fraction of systems distinguishable does \textit{not} appear to rise with planetary radius.  However, the fraction of distinguishable systems \textit{does} rise with increasing transit depth and with decreasing stellar radius.  The large fraction of systems distinguishable at higher values of TESS magnitude is due to the fact that transit depths in general are larger for dimmer TESS detections. For example, the median transit depth for systems visible from Teide Observatory that have TESS magnitude greater than 13 is 0.00989, while the median transit depth for systems brighter than TESS magnitude 13 is 0.00118. \label{fig:MuSCAT2Distinguishable}}
\end{figure*}

Figures \ref{fig:MuSCATDistinguishable} and \ref{fig:MuSCAT2Distinguishable} show a series of histograms comparing the total number of systems observable from OAC and Teide Observatory to those that would be distinguishable as BEB false positives.  Surprisingly, the fraction of systems distinguishable does \textit{not} appear to rise with planetary radius.  However, the fraction of distinguishable systems \textit{does} rise with increasing transit depth and with decreasing stellar radius.  Indeed, Figures \ref{fig:MuSCATDistinguishable} and \ref{fig:MuSCAT2Distinguishable} indicate that transit depth is the most important indicator of whether or not MuSCAT and MuSCAT2 can distinguish a true exoplanet from a BEB false positive.  In addition, panel (d) in both figures illustrates that MuSCAT and MuSCAT2 are very powerful in discriminating BEBs for the smallest host stars (mid-to-late M dwarfs).  

\begin{figure*}
\gridline{\fig{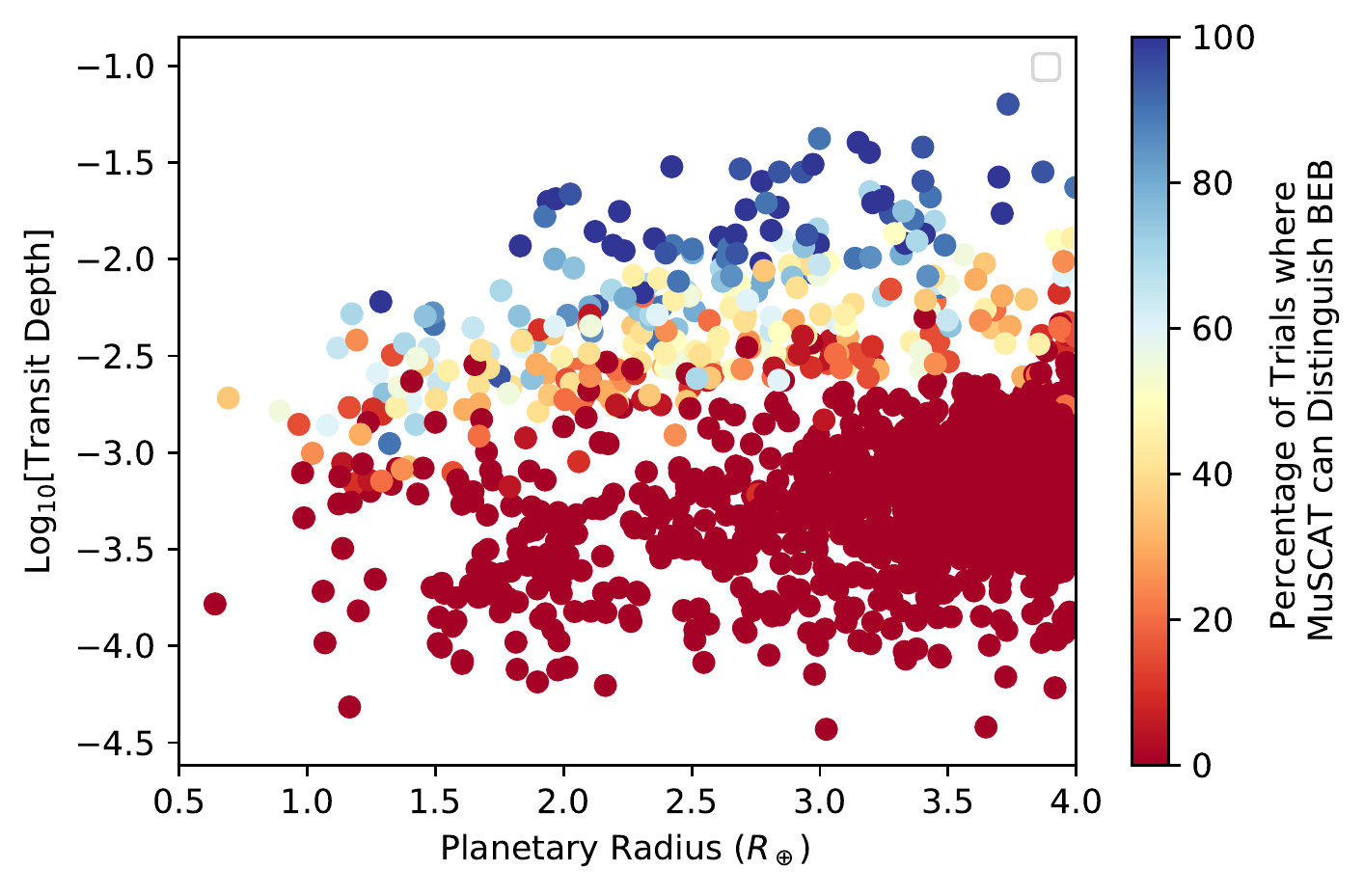}{0.5\textwidth}{(a)}
          \fig{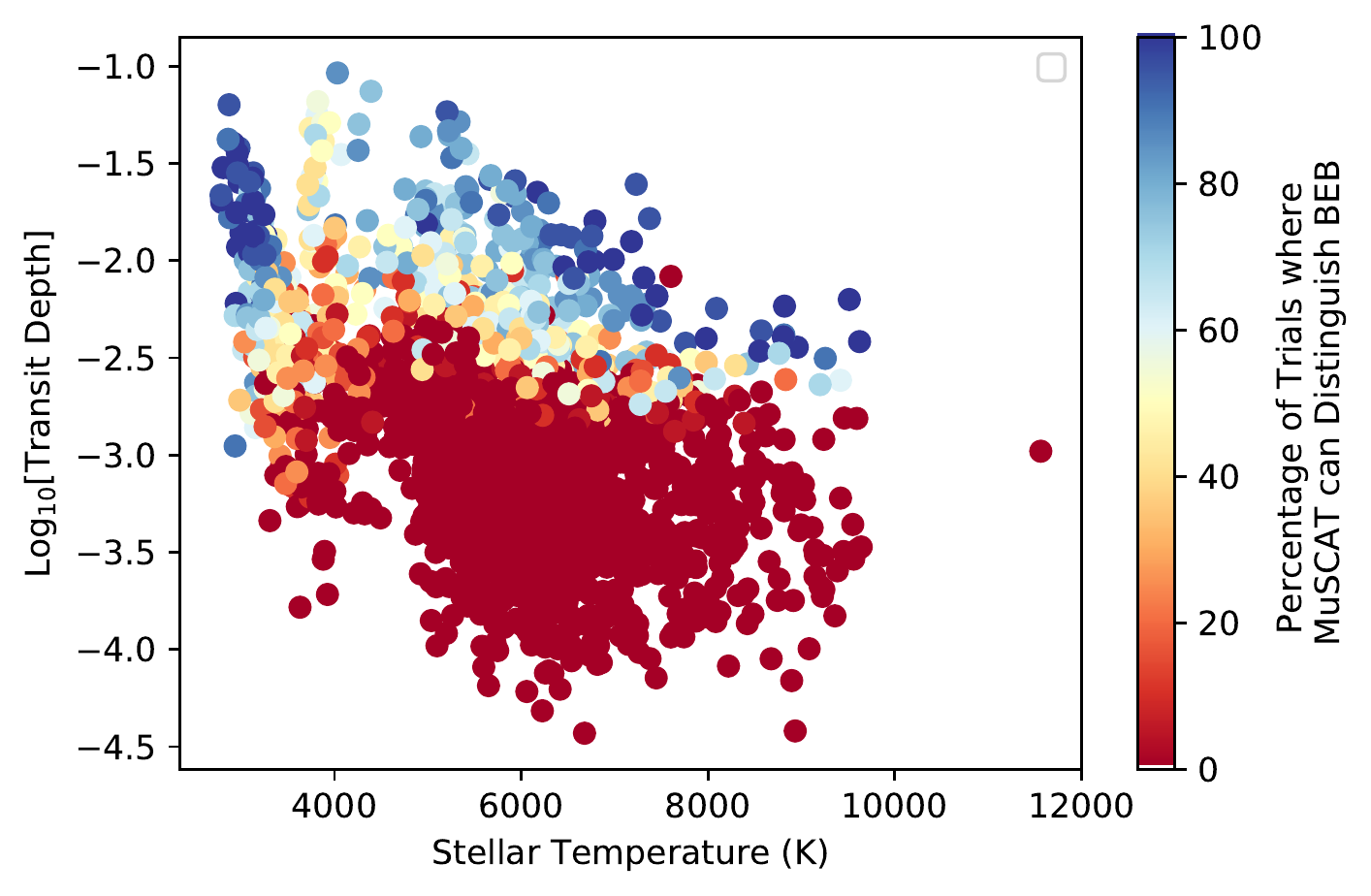}{0.5\textwidth}{(b)}}
\gridline{\fig{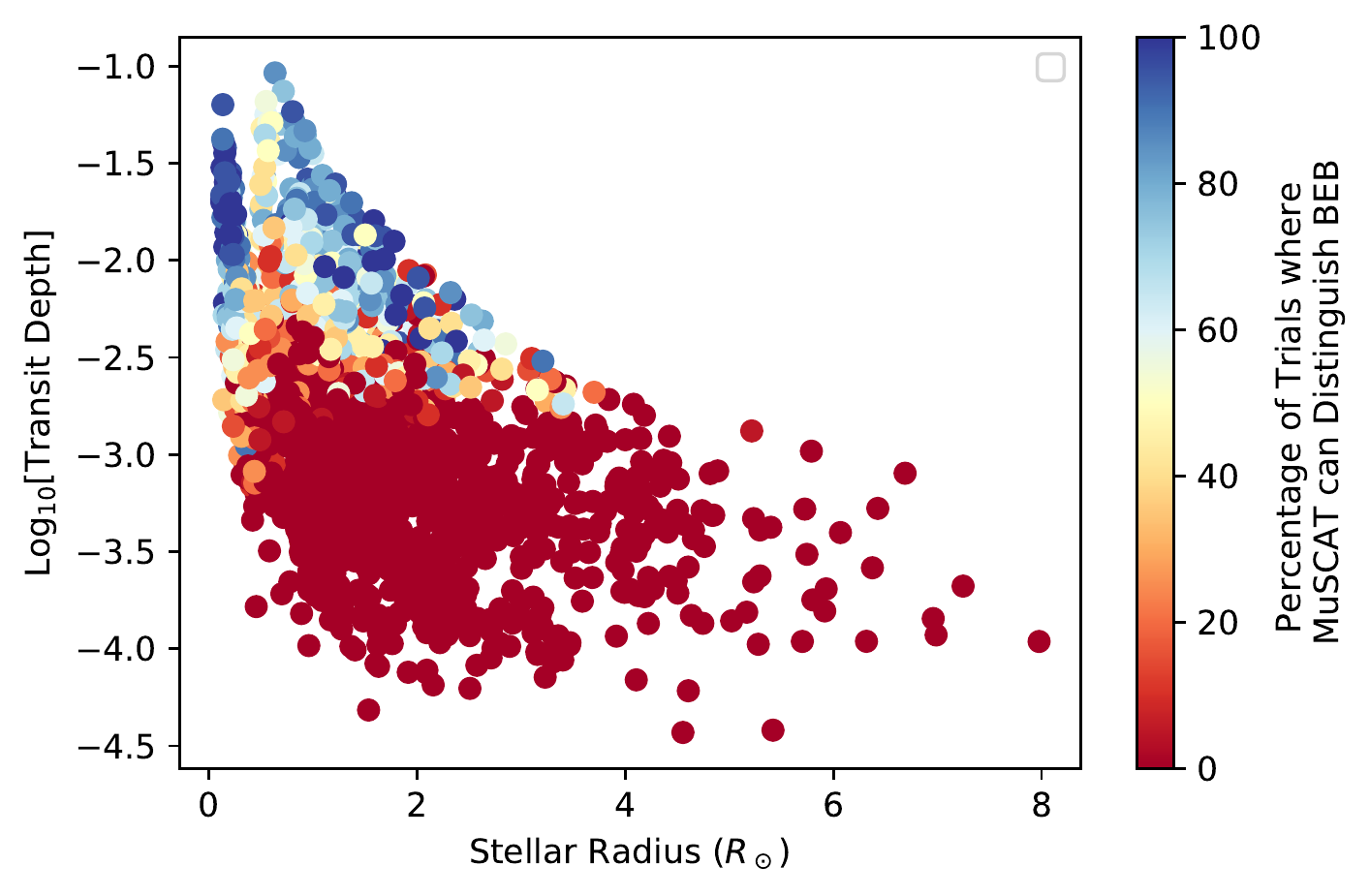}{0.5\textwidth}{(c)}
          \fig{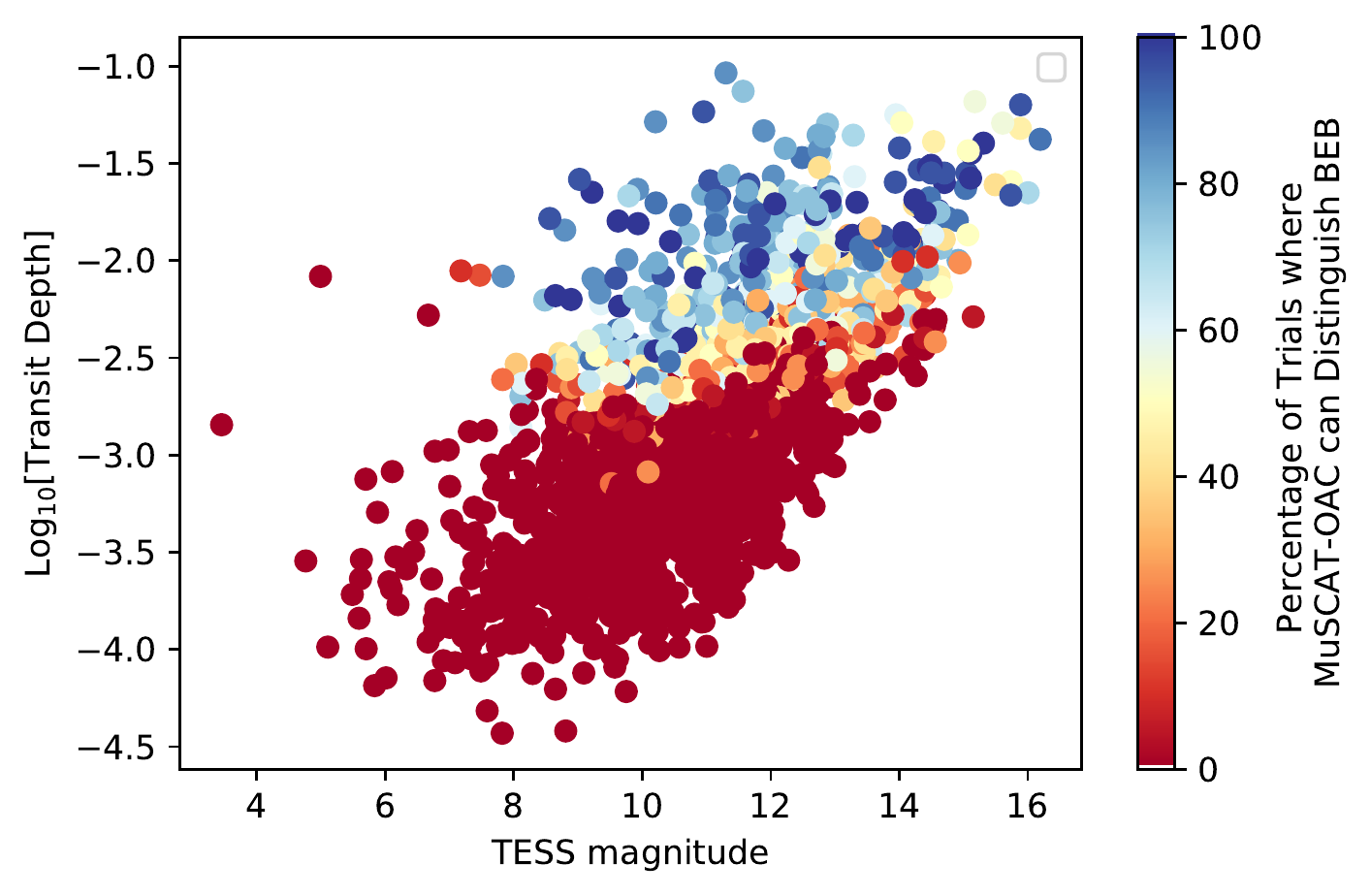}{0.5\textwidth}{(d)}}
\caption{Plots showing the percentage of trials where MuSCAT can distinguish BEB false positives from true exoplanets for the 2,575 \cite{2018ApJS..239....2B} systems visible from Okayama Astro-Complex.  Each point represents one Barclay et al. system, with the color coding indicating the percentage of trials for which that system could be distinguished as a BEB false positive.  We plot transit depth versus (a) planetary radius, (b) stellar effective temperature, (c) stellar radius, and (d) TESS-band magnitude.  MuSCAT is quite effective at distinguishing BEBs for systems with transit depths of 0.003 or greater.  For bright TESS magnitudes, MuSCAT CCD pixels saturate even with defocusing to 15 arcsec (Section \ref{sec:simtool}), and therefore we report that MuSCAT is unable to distinguish BEBs in these bright systems.  \label{fig:MuSCAT_PercentDist}}
\end{figure*}

\begin{figure*}
\gridline{\fig{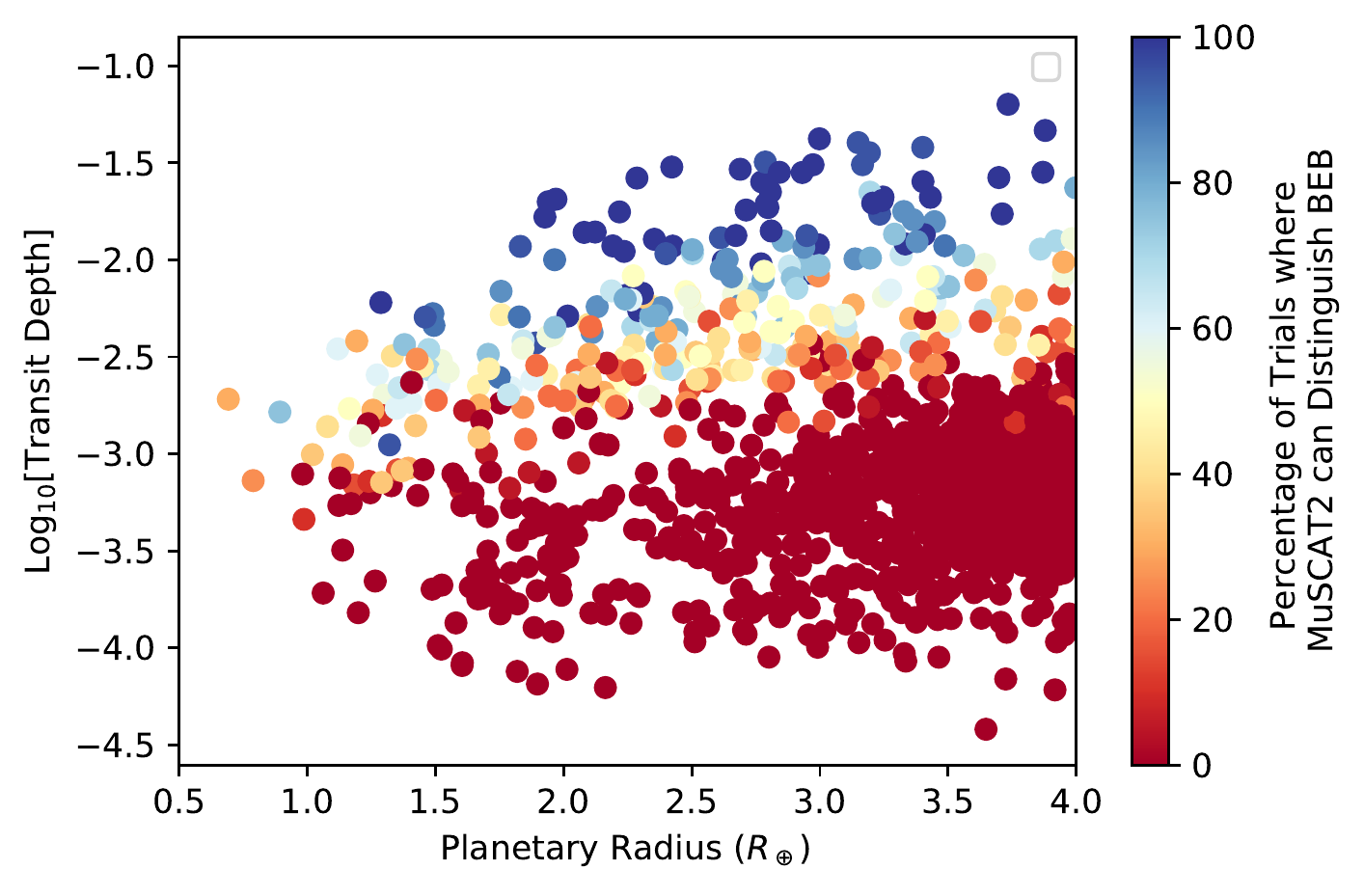}{0.5\textwidth}{(a)}
          \fig{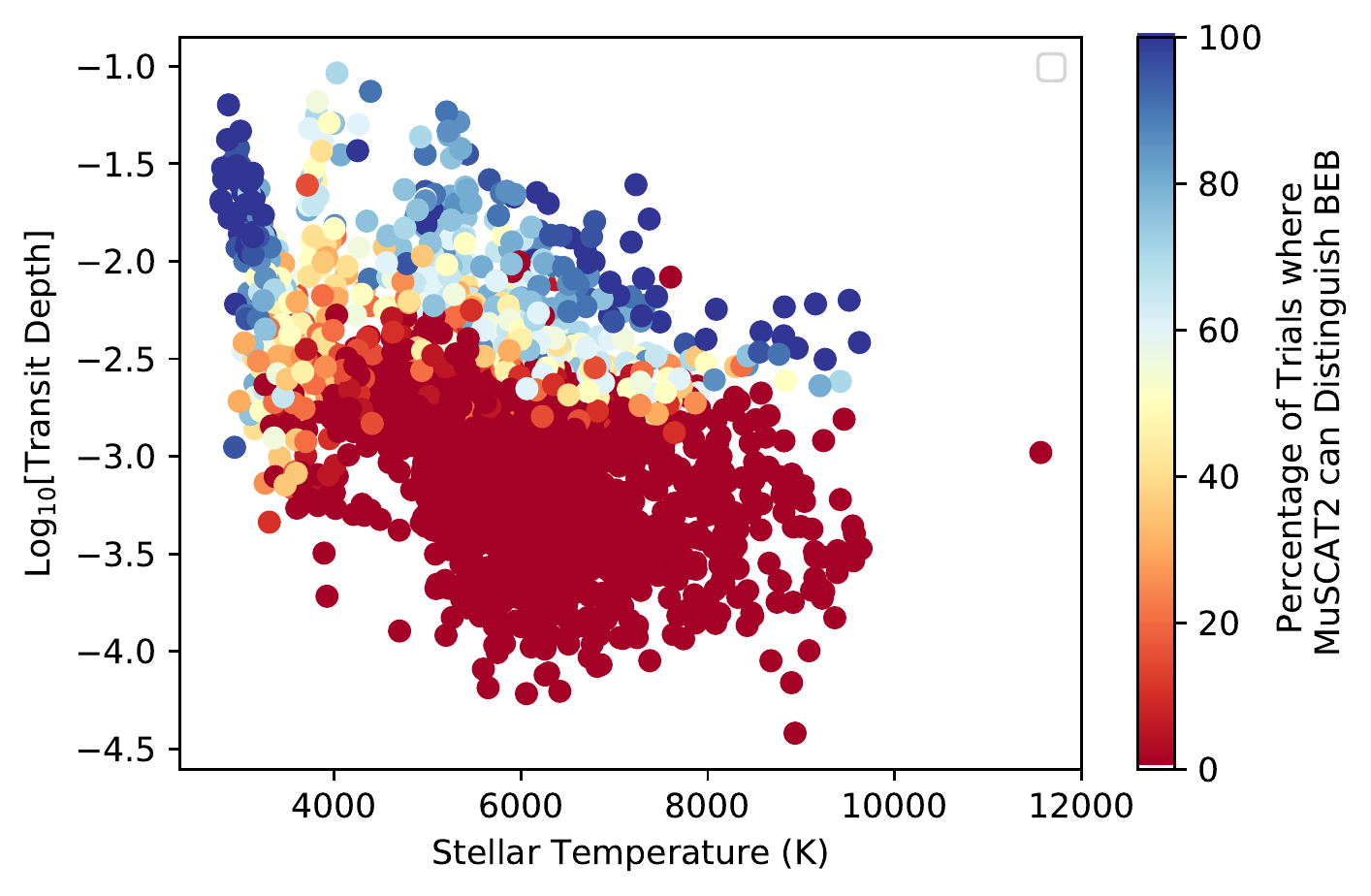}{0.5\textwidth}{(b)}}
\gridline{\fig{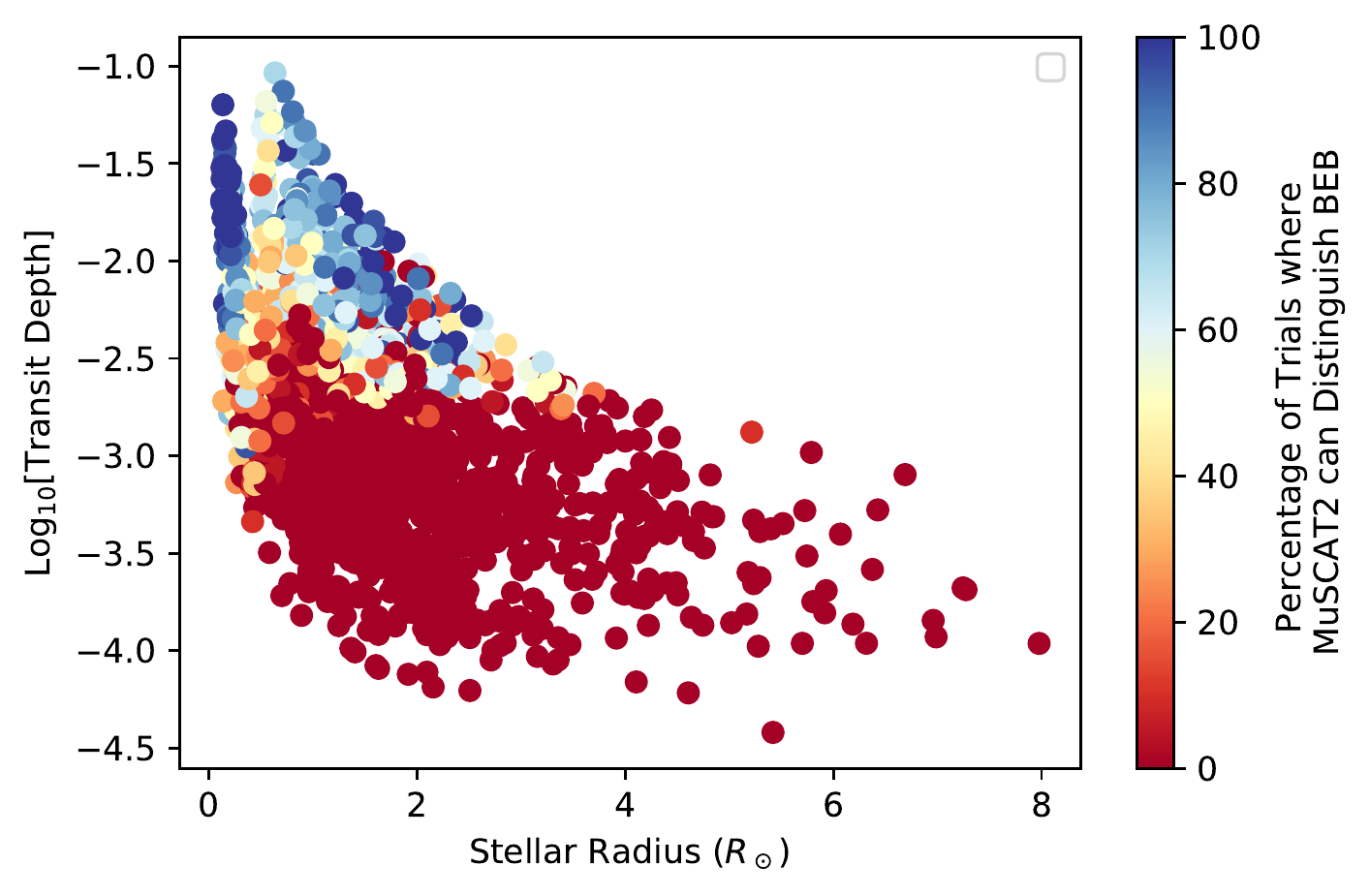}{0.5\textwidth}{(c)}
          \fig{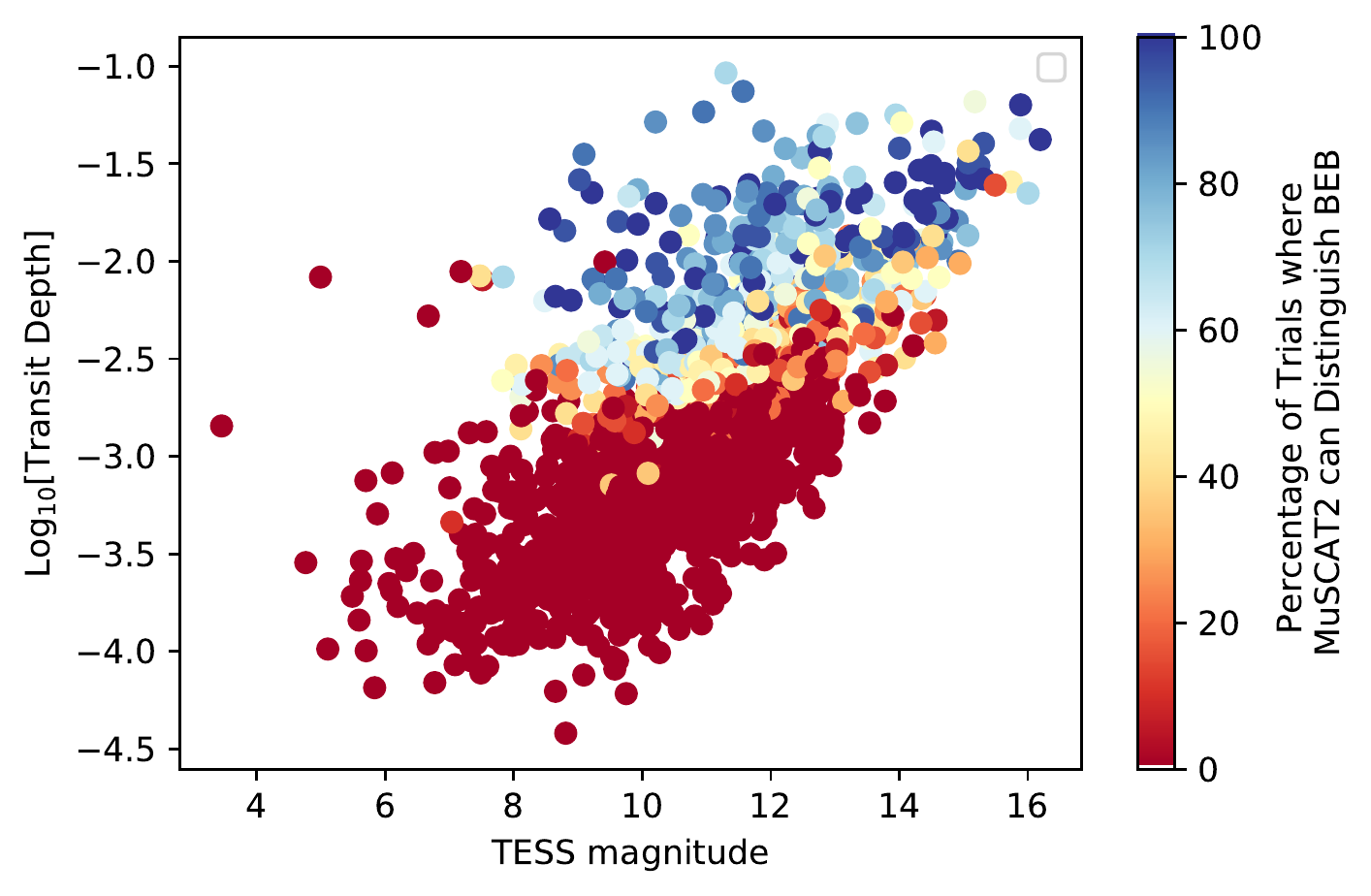}{0.5\textwidth}{(d)}}
\caption{Plots showing the percentage of trials where MuSCAT2 can distinguish BEB false positives from true exoplanets for the 2,485 \cite{2018ApJS..239....2B} systems visible from Teide Observatory.  Each point represents one Barclay et al. system, with the color coding indicating the percentage of trials for which that system could be distinguished as a BEB false positive.  We plot transit depth versus (a) planetary radius, (b) stellar effective temperature, (c) stellar radius, and (d) TESS-band magnitude.  MuSCAT2 is quite effective at distinguishing BEBs for systems with transit depths of 0.003 or greater.  For bright TESS magnitudes, MuSCAT2 CCD pixels saturate even with defocusing to 15 arcsec (Section \ref{sec:simtool}), and therefore we report that MuSCAT2 is unable to distinguish BEBs in these bright systems.   \label{fig:MuSCAT2_PercentDist}}
\end{figure*}

We examine transit depth further in Figures \ref{fig:MuSCAT_PercentDist} and \ref{fig:MuSCAT2_PercentDist}.  Here, for each \cite{2018ApJS..239....2B} system, we plot transit depth versus planetary radius, stellar effective temperature, stellar radius, and TESS-band magnitude.  The points correspond to each Barclay et al. system, and are color coded to correspond to the percentage of trials for which that system could be discriminated as a BEB false positive.  For example, dark red color coding indicates that a given system could not be distinguished for any trial as a BEB false positive.  Conversely, dark blue color coding indicates that a given system could be distinguished as a BEB false positive for 100\% of the trials.  In general, the figures confirm that transit depth is the most important characteristic in determining whether MuSCAT or MuSCAT2 can distinguish between true exoplanets and false positives.  Both instruments are quite effective in distinguishing BEBs for systems with transit depths of 0.003 or greater.

Finally, we note some of the limitations of this study.  First, our analysis assumes that we defocus the MuSCAT and MuSCAT2 instruments during observations (Section \ref{sec:simtool}), and that light from the targeted star and the BEB component stars blends together.  However, if the instruments are not defocused during observations, and if the target star and BEB are sufficiently spatially separated, then the target star and BEB can be observed separately so that the nature of the TOI is much more easily determined.  For this reason, our estimates for the number of planetary candidates that MuSCAT and MuSCAT2 can discriminate should be considered as conservative minimal estimates to the number of TOIs that the two instruments can actually validate.  In addition, we note that this study does not take into account observational factors such as weather or scheduling.  Although the capabilities of MuSCAT and MuSCAT2 appear to be similar, in reality MuSCAT2 is likely to validate many more candidates than MuSCAT.  The MuSCAT2 developers group has 162 guaranteed nights per year on MuSCAT2,\footnote{http://vivaldi.ll.iac.es/OOCC/iac-managed-telescopes/telescopio-carlos-sanchez/muscat2/} and much of that time will be devoted to TESS follow-up.  In addition, useful observing time at Teide Observatory reaches up to 78\% in the summer \citep{2002ASPC..266..454V}.

\subsection{Application to TESS Objects of Interest} \label{sec:ApplicationToTESSdata}

We applied our MuSCAT and MuSCAT2 simulation tools to TESS Objects of Interest (TOIs) posted to the Exoplanet Follow-up Observing Program for TESS (ExoFOP-TESS) website,\footnote{\url{https://exofop.ipac.caltech.edu/tess/}} to predict the probabilities that recent TOIs can be distinguished from BEB false positives using MuSCAT and MuSCAT2.  At the time ExoFOP-TESS was accessed, 995 TOIs had been uploaded to the database, spanning TESS observation sectors 1 through 12.  Since all sectors were in the Ecliptic South, we found that only 334 systems would be observable from OAC (407 systems from Teide Observatory) at some time during the year through an airmass of 2 or less.  Of the observable TOIs, the simulation tool could not be applied to 76 systems visible with MuSCAT (92 systems for MuSCAT2) because the database lacked parameters required by our routine.  For example, some systems lacked values for stellar effective temperature or log(g), which are required to select a stellar model.  Other systems lacked estimated values for planetary radius, which we use to calculate transit depth.  Below, we discuss the results of applying our routine to the 258 systems visible from OAC (315 systems from Teide Observatory), for which all required parameters were available.

\begin{figure*}
\gridline{\fig{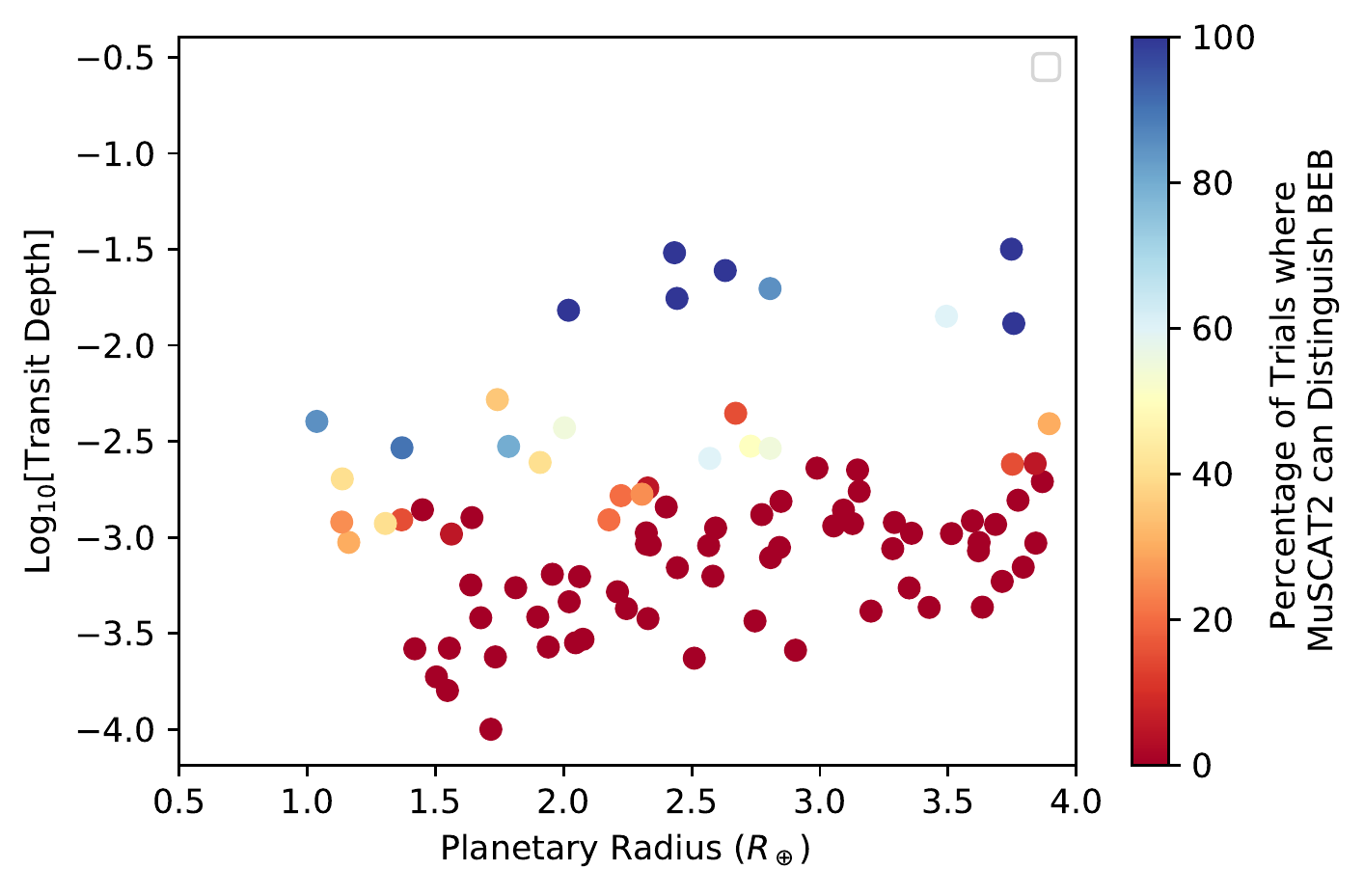}{0.5\textwidth}{(a)}
          \fig{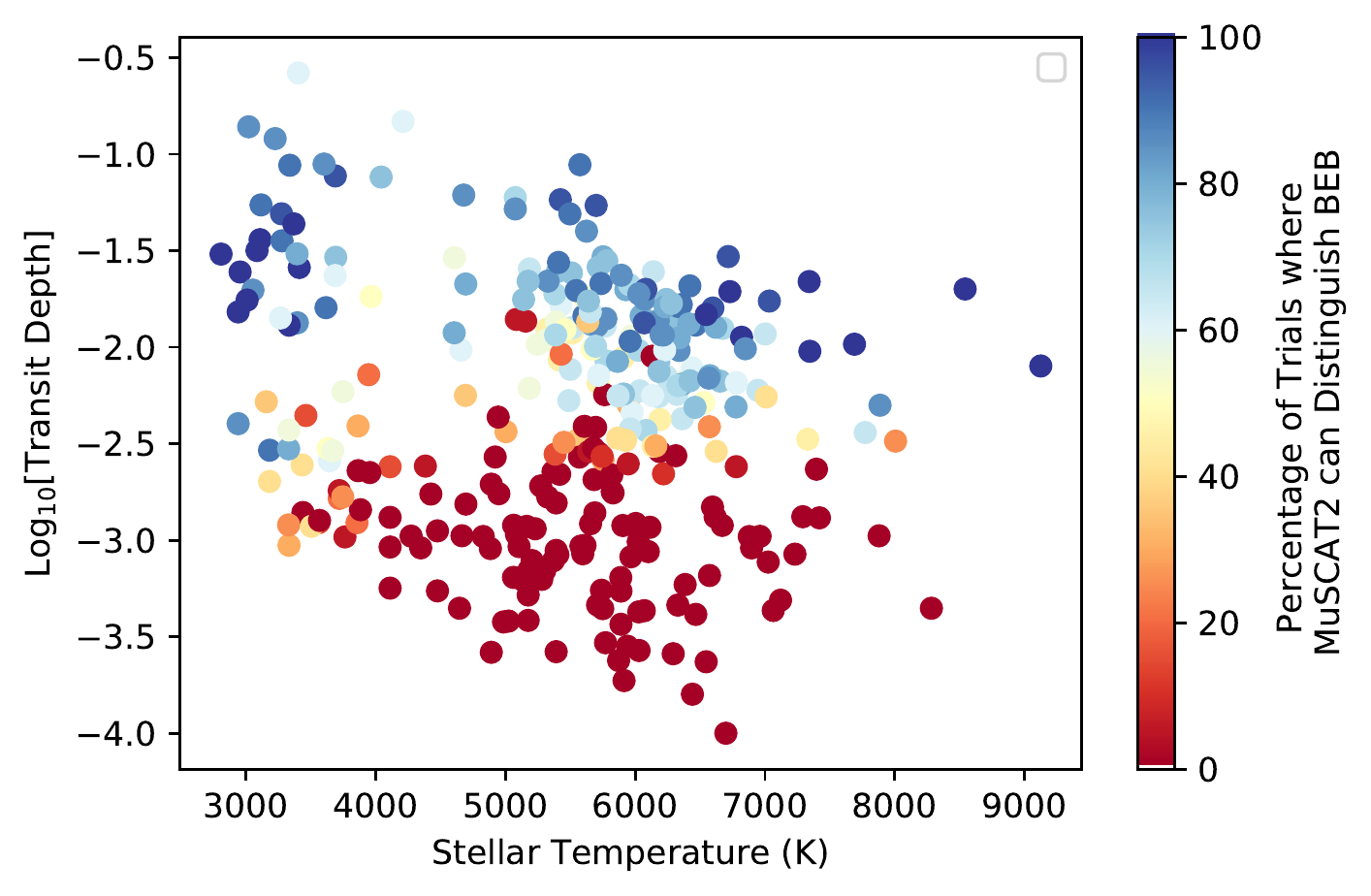}{0.5\textwidth}{(b)}}
\gridline{\fig{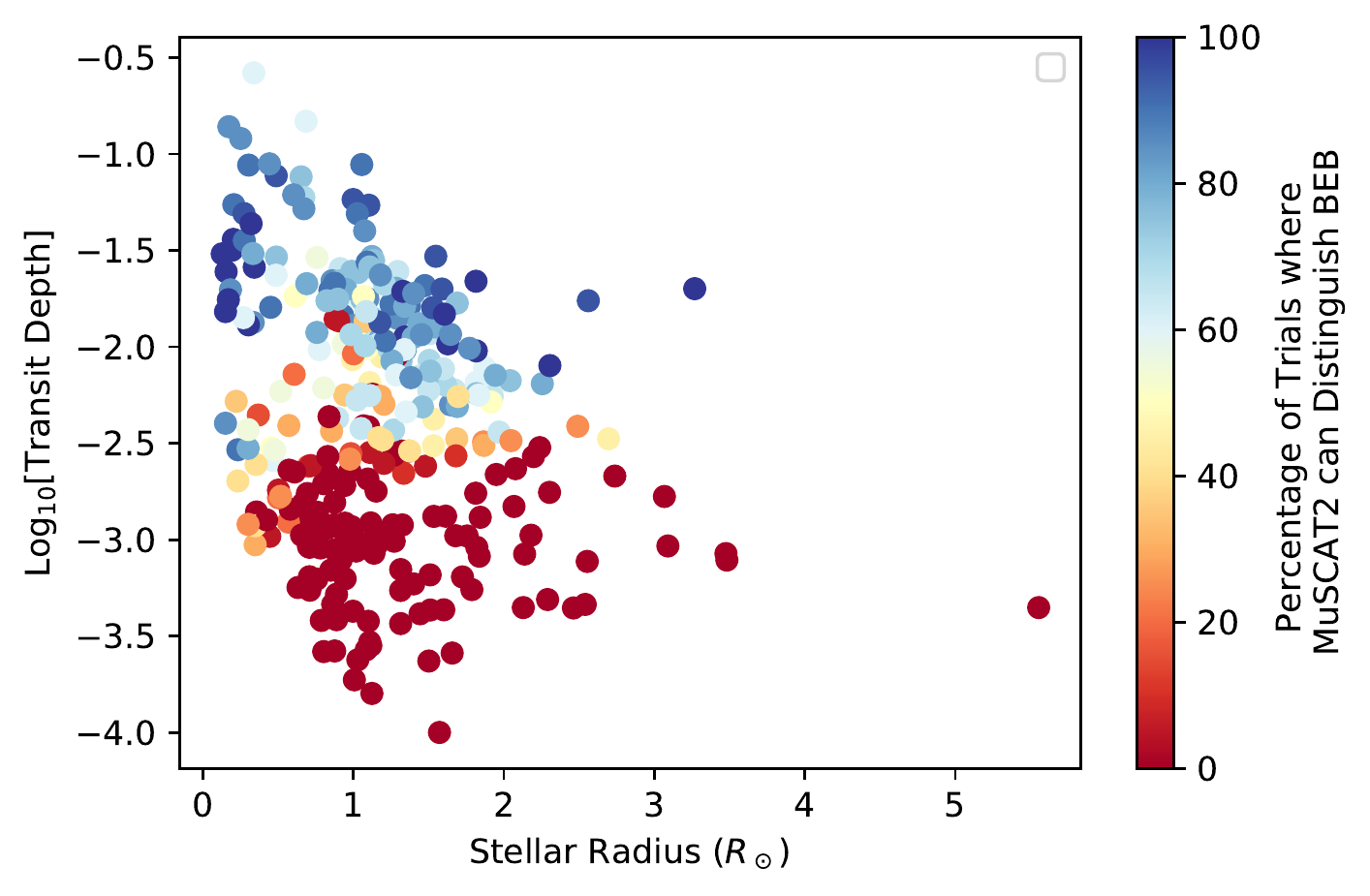}{0.5\textwidth}{(c)}
          \fig{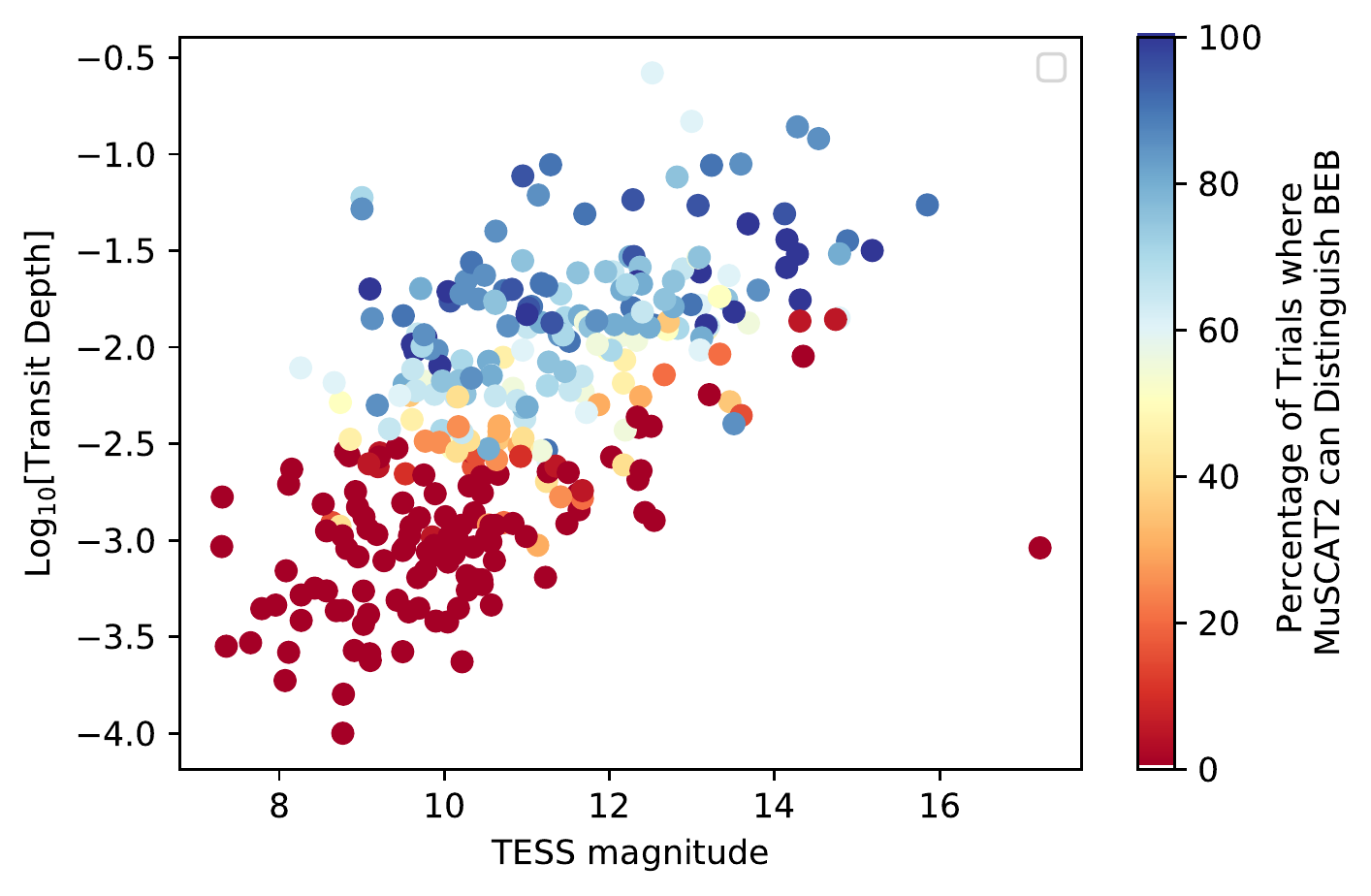}{0.5\textwidth}{(d)}}
\caption{Plots showing the percentage of trials where MuSCAT2 can distinguish BEB false positives from true exoplanets for 315 TESS Objects of Interest (TOIs) visible from Teide Observatory.  TOIs were downloaded from ExoFOP-TESS on 16 August 2019.  MuSCAT2 is quite effective at distinguishing BEBs for systems with transit depths of 0.003 or greater.  For the TOIs that we analyzed, the plots indicate that MuSCAT2 is able to distinguish BEBs at slightly smaller transit depths for smaller planetary radii (a), smaller stellar effective temperatures (b) and radii (c), and smaller TESS magnitudes (i.e., brighter systems) (d).  \label{fig:MuSCAT2_DistinguishableTOIs}}
\end{figure*}

In applying our simulation routine to the TOIs, we confirmed that MuSCAT and MuSCAT2's ability to distinguish BEB false positives depends largely upon transit depth.  We show this in Figure \ref{fig:MuSCAT2_DistinguishableTOIs}, where for MuSCAT2 we plot transit depth versus planetary radius, stellar temperature, stellar radius, and TESS magnitude, respectively, indicating with a color bar the percentage of trials where MuSCAT2 can distinguish the BEBs from true exoplanets. The plots indicate that MuSCAT2 is quite effective at distinguishing BEBs for systems with transit depths of 0.003 or greater.  In addition, for the TOIs that we analyzed, MuSCAT2 is able to distinguish BEBs at slightly smaller transit depths for smaller planetary radii, smaller stellar effective temperatures and radii, and smaller TESS magnitudes (i.e., brighter systems).  

In Tables \ref{tab:MuSCAT2TOINumDistPerTD} and \ref{tab:MuSCATTOINumDistPerTD}, we show our predictions for the number of TOIs distinguishable by MuSCAT2 and MuSCAT at transit depths greater than 0.001, 0.002, and 0.003, respectively.  The values reported confirm that both instruments are quite effective at distinguishing BEBs for systems with transit depths of 0.003 or greater, but they also show that the two instruments are quite effective for even smaller transit depths.  For example, they can discriminate over half of the systems with transit depths of 0.001 or greater. 

\begin{deluxetable*}{cccc}
\tablecaption{Predicted number of TOI systems where MuSCAT2 can distinguish true exoplanets from BEB false positives (20 trials)\label{tab:MuSCAT2TOINumDistPerTD}}
\tablewidth{0pt}
\tablehead{
\colhead{Transit} & \colhead{Number} & \colhead{Total Number} & \colhead{Percentage}  \\
\colhead{Depth (TD)} & \colhead{Distinguishable} & \colhead{of TOIs\tablenotemark{i}} & \colhead{(\%)}} 
\startdata
{TD $>$ 0.001} & 135 $\pm$ \ 5 & 259 & 52 \vspace{2mm}\\
{TD $>$ 0.002} & 133 $\pm$ \ 5 & 217 & 61 \vspace{2mm} \\
{TD $>$ 0.003} & 128 $\pm$ \ 4 & 187 & 68 \vspace{2mm} \\
\enddata
\tablenotetext{i}{\ In this column, we report the total number of systems at the indicated transit depth, out of the 315 TOIs  from TESS' Southern Ecliptic survey that are observable using MuSCAT2, for which all required parameters were available.}
\end{deluxetable*}

\begin{deluxetable*}{cccc}
\tablecaption{Predicted number of TOI systems where MuSCAT can distinguish true exoplanets from BEB false positives (20 trials)\label{tab:MuSCATTOINumDistPerTD}}
\tablewidth{0pt}
\tablehead{
\colhead{Transit} & \colhead{Number} & \colhead{Total Number} & \colhead{Percentage}  \\
\colhead{Depth (TD)} & \colhead{Distinguishable} & \colhead{of TOIs\tablenotemark{i}} & \colhead{(\%)}} 
\startdata
{TD $>$ 0.001} & 115 $\pm$ \ 5 & 210 & 55 \vspace{2mm}\\
{TD $>$ 0.002} & 114 $\pm$ \ 5 & 179 & 64 \vspace{2mm} \\
{TD $>$ 0.003} & 108 $\pm$ \ 5 & 153 & 70 \vspace{2mm} \\
\enddata
\tablenotetext{i}{\ In this column, we report the total number of systems at the indicated transit depth, out of the 258 TOIs  from TESS' Southern Ecliptic survey that are observable using MuSCAT, for which all required parameters were available.}
\end{deluxetable*}

Although not depicted here, our analysis for MuSCAT gives similar results to those shown in Figure \ref{fig:MuSCAT2_DistinguishableTOIs}.  In addition, for MuSCAT, the median transit depth for those systems that are distinguishable \textit{less than} 50\% of the time is 0.00122, while that for systems that are \textit{never} distinguishable is 0.000886.  For MuSCAT2, the median transit depth for those systems that are distinguishable \textit{less than} 50\% of the time is 0.00131, while that for systems that are \textit{never} distinguishable is 0.000941.

We can use our simulation tools to predict the probability that a given system can be distinguished as either a true exoplanet or a false positive.  Overall, for MuSCAT we found that 16 systems ($\sim$6\%) could be distinguished 100\% of the time (all 20 trials).  For MuSCAT2, 17 systems ($\sim$5\%) could be distinguished 100\% of the time.  Table \ref{tab:MuSCAT2_100PercDist} lists the 17 TOIs  always distinguished by MuSCAT2 in order of increasing stellar temperature.   For MuSCAT, thirty-five systems ($\sim$14\%) were distinguishable at least 90\% of the time, while 42 systems ($\sim$13\%) were distinguishable $\geqslant$90\% of the time with MuSCAT2.  For both instruments, approximately half of the systems (135 systems out of 258 for MuSCAT and 158 systems out of 315 for MuSCAT2) could be distinguished $\geqslant$50\% of the time.  Eighty-one systems ($\sim$31\%) could \textit{never} be distinguished by MuSCAT, and 107 systems ($\sim$34\%) could never be distinguished by MuSCAT2.    

When we average the results over all 20 trials for each instrument, we find that MuSCAT can distinguish true exoplanets from BEBs for 115 (45\%) $\pm$5 of all TOIs, while MuSCAT2 can do so for 135 (43\%) $\pm$5 of all TOIs.  For $R_{\rm pl} < 4R_\oplus$ candidates, MuSCAT can distinguish true exoplanets from BEBs for 13 (18\%) $\pm$2 of the 71 TOIs for which we have sufficient parameters to run our simulation tools, while MuSCAT2 can do so for 16 (18\%) $\pm$2 of the 90 TOIs for which we have sufficient parameters.

In July 2019, TESS began searching for transiting exoplanets in the Northern Ecliptic Hemisphere, which is composed of observation sectors 14 through 26.  MuSCAT and MuSCAT2 will be able to observe most of the TOIs discovered in these sectors.  Our results indicate that MuSCAT and MuSCAT2 will make significant contributions to the TESS Level 1 Science Requirement of measuring the masses of 50 exoplanets smaller in size than Neptune.

\begin{deluxetable*}{ccccccc}
\tablecaption{TOIs for which MuSCAT2 Can Distinguish BEBs from True Exoplanets in 100\% of Trials \label{tab:MuSCAT2_100PercDist}}
\tablewidth{0pt}
\tablehead{
\colhead{TOI\tablenotemark{i}} & \colhead{T$_{\rm eff}$} & \colhead{R$_{\rm star}$} & \colhead{T$_{\rm pl}$} & \colhead{R$_{\rm pl}$} & \colhead{TESS} & \colhead{Transit}\\
\colhead{} & \colhead{(K)} & \colhead{(R$_{\odot}$)} & \colhead{(K)} & \colhead{(R$_{\oplus}$)} & \colhead{magnitude} & \colhead{Depth}}
\startdata
227.01 & 2,808 & 0.128 & 187 & 2.43 & 14.3 & 0.0302 \vspace{2mm}\\
736.01 & 2,940 & 0.150 & 302 & 2.02 & 13.5 & 0.0151 \vspace{2mm}\\
278.01 & 2,955 & 0.154 & 781 & 2.63 & 13.1 & 0.0244 \vspace{2mm} \\
549.01 & 3,009 & 0.169 & 680 & 2.44 & 14.3 & 0.0174 \vspace{2mm} \\
543.01 & 3,085 & 0.193 & 714 & 3.75 & 15.2 & 0.0315 \vspace{2mm} \\
516.01 & 3,109 & 0.202 & 581 & 4.18 & 14.2 & 0.0358 \vspace{2mm} \\
497.01 & 3,333 & 0.302 & 492 & 3.76 & 13.2 & 0.0130 \vspace{2mm} \\
643.01 & 3,369 & 0.320 & 339 & 7.28 & 13.7 & 0.0432 \vspace{2mm} \\
538.01 & 3,411 & 0.341 & 632 & 5.98 & 14.1 & 0.0259 \vspace{2mm} \\
1050.01 & 6,548 & 1.60 & 1,619 & 21.3 & 11.0 & 0.0148 \vspace{2mm} \\
951.01 & 6,730 & 1.33 & 1,554 & 20.2 & 10.0 & 0.0193 \vspace{2mm} \\
471.01 & 6,820 & 1.34 & 1,426 & 15.5 & 9.78 & 0.0112 \vspace{2mm} \\
577.01 & 7,341 & 1.81 & 1,917 & 29.3 & 12.3 & 0.0217 \vspace{2mm} \\
508.01 & 7,346 & 1.82 & 1,722 & 19.3 & 9.64 & 0.00946 \vspace{2mm} \\
625.01 & 7,690 & 1.63 & 1,919 & 18.1 & 9.61 & 0.0103 \vspace{2mm} \\
433.01 & 8,543 & 3.27 & 3,578 & 50.4\tablenotemark{ii} & 9.10 & 0.0199 \vspace{2mm} \\
627.01 & 9,126 & 2.31 & 3,157 & 22.5 & 9.95 & 0.00794 \vspace{2mm} \\
\enddata
\tablenotetext{i}{Listed in order of increasing stellar effective temperature}
\tablenotetext{ii}{Although listed as a TOI, the calculated radius of this candidate exoplanet is larger than that of any known exoplanet, and thus the system is likely an eclipsing binary.}
\end{deluxetable*}

\section{Summary and Conclusion} \label{sec:summary}

MuSCAT and MuSCAT2 can validate hundreds of $R_{\rm pl} < 4R_\oplus$ candidate exoplanets, thus supporting the TESS team in achieving its Level 1 Science Requirement of measuring the masses of 50 exoplanets smaller in size than Neptune.  Specifically, we draw the following conclusions.

\begin{enumerate}
    \item Transit depth is the most important characteristic in determining whether or not MuSCAT and MuSCAT2 can distinguish between true exoplanets and BEB false positives.  The two instruments are most effective at distinguishing BEBs for systems with transit depths of 0.003 or greater.  
    \item We estimate that MuSCAT can distinguish BEB false positives for $\sim$17\% of all TESS discoveries, and $\sim$13\% of $R_{\rm pl} < 4R_\oplus$ discoveries.  
    \item We predict MuSCAT2 will be able to distinguish BEB false positives for $\sim$18\% of all TESS discoveries, and $\sim$15\% of $R_{\rm pl} < 4R_\oplus$ discoveries.  
    \item In analyzing actual TESS objects of interest (TOIs) from the Southern Ecliptic Hemisphere, we predict that MuSCAT can distinguish true exoplanets from BEBs for 115 (45\%) $\pm$5 of all observable TOIs, and for 13 (18\%) $\pm$2 of $R_{\rm pl} < 4R_\oplus$ observable planet candidates.
    \item In analyzing TOIs from the Southern Ecliptic Hemisphere, we predict that MuSCAT2 can distinguish true exoplanets from BEBs for 135 (43\%) $\pm$5 of all observable TOIs, and for 16 (18\%) $\pm$2 of $R_{\rm pl} < 4R_\oplus$ observable planet candidates.
    \item In analyzing TOIs from the Southern Ecliptic Hemisphere, we estimate that MuSCAT can distinguish true exoplanets from BEBs for 115 (55\%) $\pm$5 TOIs with transit depths greater than 0.001, for 114 (64\%) $\pm$5 TOIs with transit depths greater than 0.002, and for 108 (70\%) $\pm$5 TOIs with transit depths greater than 0.003. 
    \item In analyzing TOIs from the Southern Ecliptic Hemisphere, we estimate that MuSCAT2 can distinguish true exoplanets from BEBs for 135 (52\%) $\pm$5 TOIs with transit depths greater than 0.001, for 133 (61\%) $\pm$5 TOIs with transit depths greater than 0.002, and for 128 (68\%) $\pm$4 TOIs with transit depths greater than 0.003.
\end{enumerate}
Our software tools will assist TFOP working group members as they prioritize and optimize follow-up observations of TESS objects of interest. 

\acknowledgments

We thank the anonymous referee, whose helpful comments improved the quality of this paper. D.L. acknowledges support from NASA Headquarters under the NASA Earth and Space Science Fellowship (NESSF) Program - Grant NNX16AP52H.  D.L. acknowledges support from the National Science Foundation (NSF) and the Japan Society for the Promotion of Science (JSPS) under the East Asia and Pacific Summer Institutes (EAPSI) program, award number 1713804. This work is partly supported by JSPS KAKENHI Grant Numbers JP17H04574,  JP18H01265 and 18H05439, and JST PRESTO Grant Number JPMJPR1775.  This work is partly financed by the Spanish Ministry of Economics and Competitiveness through projects ESP2016-80435-C2-2-R and PGC2018-098153-B-C31.  M.T. is supported by MEXT/JSPS KAKENHI grant Nos. 18H05442, 15H02063, and 22000005.  This research has made use of the Exoplanet Follow-up Observation Program website, which is operated by the California Institute of Technology, under contract with the National Aeronautics and Space Administration under the Exoplanet Exploration Program.

%



\software{astropy \citep{2013A&A...558A..33A}, numpy \citep{2011CSE....13b..22V}, scipy \citep{2020SciPy-NMeth}, matplotlib \citep{2007CSE.....9...90H} 
          }













\end{document}